\documentclass[aps,prb,twocolumn,floatfix,footinbib,showpacs,superscriptaddress]{revtex4-1}
\usepackage{graphicx}
\usepackage{amsfonts,amsmath,amssymb,dcolumn}
\usepackage{amsthm}
\usepackage{dsfont,bm}
\usepackage{color}
\usepackage{soul} 
\usepackage{amsbsy}
\usepackage{upgreek}
\usepackage[colorlinks=true,linkcolor=blue,filecolor=blue,menucolor=blue,urlcolor=blue,citecolor=blue,anchorcolor=blue]{hyperref}%

\usepackage{mathbbol}

\usepackage{sidecap}

\newcommand*{\rttensor}[1]{\overline{\overline{#1}}}
\newcommand{\var}[1]{{\operatorname{\mathit{#1}}}}

\begin{document}

\title{Strain-Engineered Widely-Tunable Perfect Absorption Angle in Black Phosphorus from First-Principles}

\author{Mohammad Alidoust} 
\affiliation{Department of Physics, NTNU Norwegian University of Science and Technology, N-7491 Trondheim, Norway}
\author{Klaus Halterman} 
\affiliation{Michelson Lab, Physics Division, Naval Air Warfare Center, China Lake, California 93555, USA}
\author{Douxing Pan} 
\altaffiliation[Also at \,] {Beijing Institute of Nanoenergy and Nanosystems, Chinese Academy of Sciences, No. 30 Xueyuan Road, Haidian District, Beijing 100083, China}
\author{Morten Willatzen} 
\altaffiliation[Also at \,] {Beijing Institute of Nanoenergy and Nanosystems, Chinese Academy of Sciences, No. 30 Xueyuan Road, Haidian District, Beijing 100083, China}
\affiliation{Department of Photonics Engineering, Technical University of Denmark, Kongens Lyngby, DK-2800, Denmark}
\author{Jaakko Akola} 
\affiliation{Department of Physics, Norwegian University of Science
  and Technology, N-7491 Trondheim, Norway}
\affiliation{Computational Physics Laboratory, Tampere University, P.O. Box 692, FI-33014 Tampere, Finland}

\begin{abstract}
Using the density functional theory of electronic structure, we compute the anisotropic dielectric response
of bulk black phosphorus subject to strain. Employing the obtained permittivity tensor, we solve Maxwell's 
equations and study the electromagnetic response of a layered structure comprising a film of black phosphorus
stacked on a metallic substrate. Our results reveal that a small compressive or tensile strain, $\sim 4\%$, exerted either 
perpendicular or in the plane to the black phosphorus growth direction, efficiently controls the epsilon-near-zero 
response, and allows perfect absorption tuning from low-angle of the incident beam $\theta=0^\circ$ to high 
values $\theta\approx 90^\circ$ while switching the energy flow direction. Incorporating the spatially inhomogeneous strain model, we also find that for certain
thicknesses of the black phosphorus, near-perfect
absorption can be achieved through controlled variations of the in-plane strain.
These findings can serve as guidelines 
for designing largely tunable perfect electromagnetic wave absorber devices.    
\end{abstract} 

\date{\today}

\maketitle

\section{introduction}
Bulk black phosphorus (BP) is an anisotropic semiconductor with two types of chemical bonding. Along 
two principal crystal directions, the phosphorus atoms form covalent bonds with a puckered honeycomb 
arrangement, whereas in the third direction atoms interact relatively weakly through van der Waals forces. \cite{wei,fang,A.S.Rodin,X.Peng,Voon1,Voon2}
The latter bonding results in a layered configuration consisting of two-dimensional phosphorus sheets. 
These weakly interacting two-dimensional layers provide a unique opportunity to create different orderings 
of two-dimensional layers with extremely low-cost and simple operations, including   
displacement and twist. \cite{wei,fang,Z.Zhang,S.Das,Z.Qin,AlidoustBP1,AlidoustBP2,
AlidoustBP3,Y.Ren,D.Odkhuu,W.Li,Doux1,Doux2}
The ordering and number of layers, as well as their deformation, can effectively control the 
electronic properties of BP-based devices. For instance, the band gap of BP is highly 
sensitive to the number of BP layers so that a monolayer of BP possesses the largest band gap which  
decreases by adding more layers. The application of strain into the plane of two-dimensional layers can 
cause a number of nontrivial phenomena, including manipulation of the band structure, and 
consequently, electronic response.  

Another significant property of BP is its ability to absorb electromagnetic (EM)
waves over a broad range of wavelengths, from visible to infrared. \cite{S.Das,N.Feng,Z.Qin,J.Wang,F.Xiong,D.Li,T.Guo,D.David,T.Liu,S.Xiao,S.Zhang,
P.T.T.Le,D.Dong,N.Feng,Y.Huang,W.Shen,D.Q.Khoa,Y.M.Qing,Q.Hong,C.Fang,X.Wang,
J.Wang2,F.Xiong,D.Li} Furthermore, the absorption of EM waves by BP can be significantly 
enhanced by creating  layered geometries that generate interference phenomena. These structures  
can offer advancements in photodetectors and field effect transistors. \cite{M.Engel,J.Wang,J.Wang2,H.Wang} 
One of the main challenges in developing modern optical devices is the inherent optical loss that can  
adversely impact favorable phenomena such as elastic scattering and the transport of optical information. 
Recently, it was shown that by incorporating detailed electronic band structure effects into the light scattering 
rate, one will be able to describe the optical loss properties of a material more accurately. \cite{gp3}
 
Some recent design approaches to control the loss in absorbers involve the use of
epsilon-near-zero (ENZ)-based metamaterials,\cite{enghetta} where the EM response  
is described by a permittivity tensor $\rttensor{\varepsilon}$ with at least one component
whose real part becomes vanishingly small over certain frequencies. A number of ENZ-based 
architectures have been fabricated, including sub-wavelength dielectric coatings with ENZ 
regions that control the resonant coupling of light \cite{gal} and the propagation of a transverse 
magnetic optical beam through a sub-wavelength slit. These experiments demonstrated a transmission 
enhancement when the semiconductor substrate was tuned to its ENZ frequency.\cite{adams} Previous ENZ-based absorbers often exploit resonance or interference effects that arise from 
the large electric-field enhancement and extreme values of the propagation vectors in the ENZ 
medium. By placing a metal in contact with a material exhibiting ENZ response, the reflected 
waves from the metal can interfere via coherent perfect absorption, \cite{feng} whereby the 
incident EM beam is perfectly coupled to the structure.

In this paper, we address how the application of strain in various directions can be utilized as an 
efficient experimental knob that controls coherent perfect absorption in a bulk two-dimensional material 
stacked on a metallic substrate. 
By performing first-principles calculations, we obtain the anisotropic dielectric response of bulk BP 
subject to compressive and tensile strain. Incorporating the permittivity tensor from first-principles 
calculations within Maxwell's equations, we show that an electromagnetic wave incident on a 
semi-infinite layered BP/metallic stack (shown in Fig.~\ref{diagramxz}), can be perfectly absorbed by 
a strain-controlled ENZ response mechanism. By varying the strain, interference effects can be tuned 
to achieve perfect absorption over a wide range of incident beam angles $\theta_P$, ranging from  
near-grazing incidence  ($\theta_P\approx 0^\circ$) to near-normal incidence ($\theta_P\approx 90^\circ$). 
Also, we show that the application of a low strain value, $\approx 4\%$, results in energy flow reversal 
by $180^\circ$ within the BP region. 
Furthermore, we find that when strain is nonuniform along the direction normal to the BP layer, 
 near perfect absorption can arise 
through controlled variations in the in-plane strain.
 
The paper is organized as follows. In Sec. \ref{method}, we summarize the theoretical framework used 
to describe the proposed EM perfect absorber from atomistic-scale. In Sec. \ref{results}, we present the 
main findings and characterize how BP-based EM perfect absorber works, including thickness, angle 
of incident EM beam, and the associated ENZ response. Additional information and discussions are 
presented in Appendices \ref{appendix1} and \ref{appendix2}. Finally, we give concluding remarks in 
Sec. \ref{conclusions}.
 
\section{method and approach} \label{method}  
The dielectric function $\tilde{\varepsilon}$ of a system can be defined by the response to an external electric 
field $\rm {\bf E}_{ext}$
\begin{equation}
{\rm {\bf D}}({\bm r};\omega)=\int d{\bm r} \tilde{\varepsilon} ({\bm r}-{\bm r}';\omega){\rm {\bf E}}_{ext}({\bm r}';\omega),
\end{equation}
where ${\bm r}$ is location, $\omega$ is the frequency of external field, and ${\rm {\bf D}}$ is the total 
electric field. If the external field is sufficiently weak (compared with the internal electric fields produced 
by charge density imbalance and ion interaction), the location and time variation of charge density 
$\delta n({\bm r};t)$ can be approximated as a linearly dependent response to the external field

\begin{figure}[t!] 
\centering
\includegraphics[width=8.5cm]{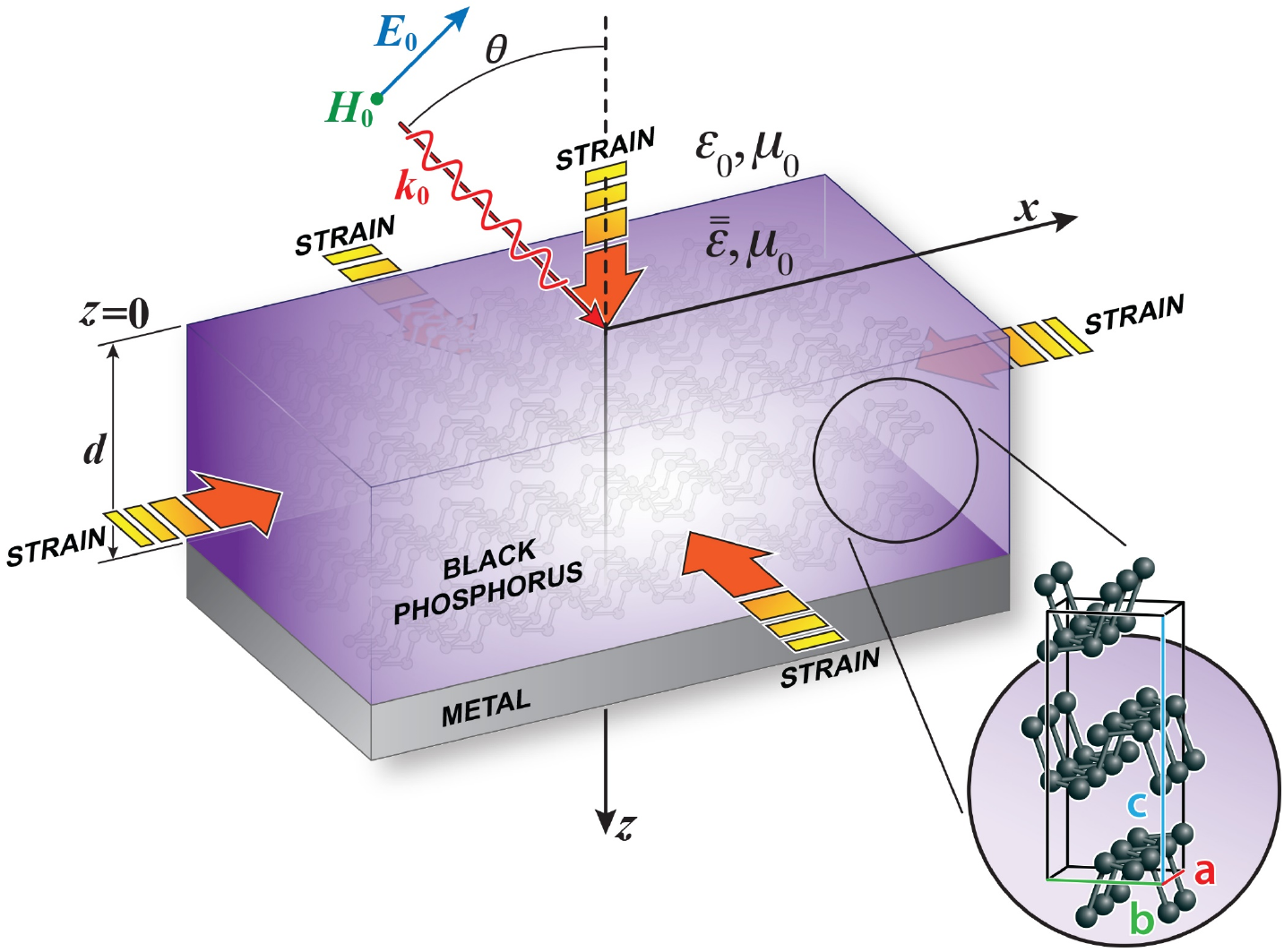} 
\caption{(Color online).  Schematic configuration involving a slab of bulk black phosphorus of thickness $d$ adjacent to a metallic substrate. The black phosphorus layer is exposed to a 
transverse magnetic field 
from the vacuum region where the incident electric field is polarized in the $\var{x-z}$ plane, and the 
magnetic field is polarized along $y$. The incident wavevector ${\bm k_0}$ makes an angle $\theta$ with 
the $z$ axis. The tensile or compressive strains are applied in the $\var{x-y}$ plane and along the $z$ 
direction (note that the arrows marking applied strain can be reversed corresponding to compressive and 
tensile strain, respectively). The crystallography principal directions are set as follows: 
$x\equiv a$, $y\equiv b$, and $z\equiv c$. }
\label{diagramxz}
\end{figure}    

\begin{equation}\label{chi}
\begin{aligned}
\delta n({\bm r};t) = &\int dt' \int d{\bm r}' \chi({\bm r},{\bm r}',t,t') \delta {\rm {\bf E}}_{ext}({\bm r}',t'),\\
&\chi({\bm r},{\bm r}',t,t')=\frac{\delta n({\bm r};t)}{\delta {\rm {\bf E}}_{ext}({\bm r}',t')}.
\end{aligned}
\end{equation} 
Within the context of the density functional theory (DFT), a weak variation in the effective potential $\rm V_{eff}$ 
of the Kohn-Sham equations results in the variation of charge density
\begin{equation}\label{ks}
\begin{aligned}
\delta n({\bm r};t) = &\int dt' \int d{\bm r}' {\cal K}({\bm r},{\bm r}',t,t') \delta {\rm V_{eff}}({\bm r}',t'),\\
&{\cal K}({\bm r},{\bm r}',t,t')=\frac{\delta n({\bm r};t)}{\delta {\rm V_{eff}}({\bm r}',t')}.
\end{aligned}
\end{equation} 
By equating Eqs. (\ref{chi}) and (\ref{ks}), one can evaluate the dielectric response function $\chi$ for an 
interacting electron system using the Kohn-Sham response function $\cal K$ from atomistic-scale DFT
simulations. \cite{AB_dielect} A well-known approximation that highly simplifies the calculations is the 
random phase approximation (RPA), neglecting the exchange-correlation contribution once the
ground-state electron density has been computed. \cite{PN} 

In this work, the atomistic-scale calculations of the dielectric response were performed in the framework 
of DFT using the $\rm GPAW$ program package, which utilizes on the projector-augmented wave (PAW) 
method for the interaction between valence electrons and ion cores. \cite{gp1,gp2,gp3} The 
gradient-corrected functional by Perdew-Burke-Ernzerhof (PBE) was employed for the exchange-correlation 
energy (electronic band structure and density of states). We have used a relatively 
high value, i.e., $6.0$~${\bf k}$-points per $\text{\AA}^{-1}$ for the ${\bf k}$-point density in order to grid ${\bf k}$-space on the basis of the 
Monkhorst-Pack scheme. The plane-wave cut-off for the kinetic energy was set to $800$ eV and 60 unoccupied electronic bands were chosen for the unit cell of eight atoms with a convergence on the first 50 bands to avoid any artificial effects that strain may induce around the Fermi energy. Correspondingly, the width of the Fermi-Dirac distribution was set to $0.01$ eV. RPA is used for the density response function, and the integrations to obtain the dielectric response are 
performed by the linear tetrahedron interpolation scheme as implemented in $\rm GPAW$. \cite{gp3} 
A small imaginary part was added to the frequencies with $\delta\omega=0.01$ eV. 

To simulate BP under strain, we introduce the strain parameters $\epsilon_{ii}$, (for $i=x,y,z$)
corresponding to the normalized percentage of uniform shrinkage with respect to relaxed unit cell. 
We define,
$a=\epsilon_{xx}  a_0$,
$b = \epsilon_{yy}  b_0$, and
$c =\epsilon_{zz} c_0$,
where $a$, $b$, and $c$ are the three strained unit cell axis lengths, and the unstrained unit cell axis 
lengths are $a_0$, $b_0$, and $c_0$. The exact values of these parameters and the location of 
phosphorus atoms in the unstrained unit cell are summarized in Table \ref{abctab} of Appendix 
\ref{appendix2}. We also present the electronic band structure and density of states for the unstrained 
BP as well as its strained forms in Appendix \ref{appendix2}. 
Hence, in this notation, $\epsilon_{xx}=\epsilon_{yy}=\epsilon_{zz}=1.0$ corresponds to zero strain, and, 
e.g., a compressive or tensile strain of $\mp 10\%$ in the $z$-direction is denoted by $\epsilon_{zz}=0.9$ 
and $\epsilon_{zz}=1.1$, respectively. Note that both unit cell parameters $a,b,c$ and correspondingly, 
the location of atoms are renormalized in the presence of strain. The permittivity tensor $\rttensor{\varepsilon}$ 
takes the following biaxial form that is valid for BP belonging to the $D_{2h}$ point group:
\begin{equation}
\rttensor{\varepsilon}_n = \varepsilon_{nx} \hat{{\bm x}} \hat{{\bm x}}
+\varepsilon_{ny} \hat{{\bm y}} \hat{{\bm y}}
+\varepsilon_{nz} \hat{{\bm z}} \hat{{\bm z}},
\end{equation}
where $n$ denotes either the vacuum region ($n=0$) or BP region ($n=1$). In general, from
symmetry considerations, the permeability tensor $\rttensor{\mu}$ is also biaxial; however for non-magnetic BP,
$\rttensor{\mu}=\mu_0 \rttensor{I}$. Note that, we make use of symbols ``$\epsilon_{xx,yy,zz}$'' for the strain whereas ``$\varepsilon_{nx,ny,nz}$'' are used for denoting the dielectric response.

We next  demonstrate how BP structure in the low-permittivity regime can 
exhibit perfect absorption of EM waves over a broad range of incident angles
and system parameters, thus revealing a practical platform for the control of EM radiation.
We investigate the reflection and absorption of EM waves from the layered configuration shown in 
Fig.~\ref{diagramxz}, which comprises a planar BP material adjacent to a metallic substrate with 
perfect conductivity (PEC). The electric field of the incident wave is polarized in the $\var{x-z}$ plane,
so that only the permittivity components $\varepsilon_{1x}$ and $\varepsilon_{1z}$ participate  in
the overall EM response. The plane wave is incident from the vacuum region with wavevector 
${\bm k}_0$  in the $\var{x-z}$ plane: ${\bm k}_0=\hat{\bm x}k_{0x} + \hat{\bm z} k_{0z}$. 
Since there are no off-diagonal components, the TM (transverse magnetic) and TE (transverse electric)  
modes are decoupled.

The incident electric and magnetic fields  thus have the following forms
\begin{subequations}\label{EHfields}
\begin{align}
&{\bm E} =(E_{x0} \hat{\bm x}+E_{z0}\hat{\bm z}) e^{i(k_{0x} x +k_{0z} z-\omega t)},\\
&{\bm H} =H_{y0} \hat{\bm y} e^{i(k_{0x} x +k_{0z} z-\omega t)}. 
\end{align}
\end{subequations}
Due to continuity in the transverse electric field, $k_{0x}$ is invariant across the interface
with $k_{0 x} = k_0 \sin\theta$,  $k_{0 z} = k_0 \cos \theta$, and $k_0=\omega/c$.
For both the vacuum and BP regions, we  implement   
Maxwell's equations for time harmonic fields ($e^{-i \omega t}$),
\begin{subequations}\label{mxwl_homo}
\begin{align}
&{\bm \nabla} \times {\bm E}_j = i \omega \mu_0 {\bm H}_j, \\
&{\bm  \nabla} \times {\bm H}_j = -i\omega {\bm D}_j,
\end{align}
\end{subequations}
where $j=0$ or $1$ to identify either the vacuum or BP regions, respectively. 
When expressing  the EM fields in BP as  plane waves, 
the propagation vector there, ${\bm k}_1$, replaces the spatial derivatives,
transforming  Maxwell's equations into the forms, ${\bm k}_1 \times {\bm E}_1= \omega \mu_0 {\bm H}_1$ 
and ${\bm k}_1 \times {\bm H}_1 = -\omega {\rttensor{\varepsilon}_1} \varepsilon_0  {\bm E}_1$. 
These two equations together result in the following  expression for the ${\bm E}_1$ 
field in $\bm k$-space:
${\bm k}_1 \times ({\bm k}_1 \times {\bm E}_1) = -k_0^2  {\rttensor{\varepsilon}_1}  {\bm E}_1$. Upon  using the identity  
${\bm k}_1 \times ({\bm k}_1 \times {\bm E}_1) = {\bm k}_1({\bm k}_1 \cdot {\bm E}_1) - k_1^2 {\bm E}_1 $, one finds the dispersion equation for the BP region:
\begin{align} \label{disp2p}
(k_{0x}^2-\varepsilon_{1y} k_0^2+k_{1z}^2)(\varepsilon_{1x}\varepsilon_{1z} k_0^2-\varepsilon_{1x} k_{0x}^2 -\varepsilon_{1z} k_{1z}^2)=0.
\end{align}
Solving for the roots in Eq.~(\ref{disp2p})  results in two types of solutions for $k_{1z}$.
If there is a $y$ component to the electric field, then we have TE modes with 
$k_{1z} = \pm\sqrt{\varepsilon_{1y} k_0^2-k_{0x}^2}$.
For the case of interest, the electric field is polarized in the $x$ and $z$ directions
(TM modes) with the following wavevectors for each region:
\begin{align} \label{kz1}
 k_{1z} & = \pm\sqrt{\varepsilon_{1x}\left(k_0^2-\frac{k_{0x}^2}{\varepsilon_{1z}}\right)}, \quad k_{0z} = \pm \sqrt{k_0^2-k_{0x}^2}.
\end{align}
Thus it is clear that due to the initial TM polarization state, only $\varepsilon_{1x}$ and $\varepsilon_{1z}$
contribute to the EM response of BP.

For the  configuration shown in Fig.~\ref{diagramxz}, where the $\var{x-y}$ plane is translationally invariant and the thickness along the $z$-axis is finite, the magnetic field 
component in the vacuum region, ${\bm H}_0$, is written in terms of incident and reflected 
waves: $H_{y0} = (e^{i k_{0z} z} + r e^{-i k_{0z} z})e^{i k_{0x} x}$, where $r$ is the reflection coefficient. From the magnetic field component, we can use 
Eqs.~(\ref{mxwl_homo}) to easily deduce the electric field components. For BP region, 
the general form of the ${\bm E}$ field is a linear combination of waves with wavevectors 
given in Eq.~(\ref{kz1}): $E_{x1}=(a_1 e^{i k_{1z} z}+a_2e^{-i k_{1z} z})e^{i k_{0x} x}$. To construct the remaining $\bm E$ and $\bm H$ fields we  use Maxwell's equations
to get the following relations
\begin{subequations}
\begin{align}
&\partial_z H_{y1}=i\omega \varepsilon_0\varepsilon_{1x} E_{x1},\\
&k_{0x} H_{y1}=-\omega \varepsilon_0 \varepsilon_{1z} E_{z1},\\
&\partial_z E_{x1}-ik_{0x} E_{z1}=i\omega\mu_0 H_{y1}. 
\end{align}
\end{subequations}
Upon matching the tangential electric and magnetic fields at the vacuum/BP interface, and 
using the boundary conditions of vanishing tangential electric fields at the metallic ground plane,
it is straightforward to determine the unknown coefficients $a_1$, $a_2$, and $r$. 

We assume that the metallic substrate is perfectly conducting. For example, silver at a frequency 
of $20\,{\rm eV}$ and corresponding skin depth $\sim\, 1\times 10^{-3}\, \upmu{\rm m}$ behaves
as a nearly perfect reflector \cite{feng}. At high frequencies, 
the metallic substrate may no longer serve as an effective reflector, 
which in turn diminishes 
the processes involved in coherent perfect absorption.
Thus,
the device should be designed
to operate below the  plasma frequency of the chosen metal. The reflection coefficient $r$ is found to be,
\begin{align} \label{arrr}
r=\frac{2 \varepsilon_{1z} k_{0z} k_{1z} \cos(k_{1z}d)}
{\varepsilon_{1z} k_{0z} k_{1z} \cos(k_{1z}d) - i (\varepsilon_{1z} k_0^2-k_{0x}^2) \sin(k_{1z}d)}-1. 
\end{align}
The coefficients $a_1$ and $a_2$ are simply related: $a_1/a_2 = -e^{-2 i k_{1z} d}$, where
\begin{align}
a_1= \frac{{\eta}_0 k_{0z}}{k_0\cos(k_{1z}d)} \frac{  e^{-i k_{1z} d}}
  {\frac{\varepsilon_{1z} k_{0z} k_{1z}}{\varepsilon_{1z} k_0^2-k_{0x}^2} -i\tan(k_{1z}d)}
\end{align}
and ${\eta}_0=\sqrt{\mu_0/\varepsilon_0}$ is the impedance of free space. 
The reflection coefficient has the  property that $r(k_{0z}) = r^{-1}(-k_{0z})$,
which simplifies the solution process later since finding a complex pole (related to $-k_{0z}$), 
is equivalent to finding the perfect absorption modes for $k_{0z}$ \cite{pole1,pole2}. 

In determining the absorptance $A$ of the black phosphorus system, we consider energy conservation,
and implement the time-averaged Poynting vector ${\bm S}$, given by
${\bm S}=\frac{1}{2}\Re\{{\bm E}\times{\bm H}^*\}$. Considering the component in the direction perpendicular to the interfaces (the $z$ direction), 
and inserting the electric and magnetic fields calculated for the vacuum region, we find
\begin{equation}\label{Absorb}
A= 1-\left | r \right |^2.
\end{equation}
Here $A$ is defined as $S_{z}/S_0$, in which $S_0\equiv k_{0z}/(2 \varepsilon_0 \omega)$
is the time-averaged Poynting vector for a plane wave traveling in the $z$ direction.
When discussing the direction of energy flow, it is insightful to consider the angle $\theta_S$ 
that the Poynting vector makes at BP/vacuum interface. We thus consider 
$\tan \theta_{S} = \Re(S_{x})/\Re(S_{z})$. Inserting the calculated electric and magnetic fields, 
we find the following general relationships: For the direction of energy flow in the vacuum region, 
we have,
\begin{align} \label{vrel}
\tan \theta  = \frac{1-|r|^2}{|1+r|^2} \tan \theta_{S}.
\end{align}
Here $r$ is a function of the permittivities $\varepsilon_{1x}$ and $\varepsilon_{1z}$, incident angle 
$\theta$, frequency $\omega$, and thickness $d$ (see Eq.~(\ref{arrr})). 
Consequently, if there is no reflected wave ($r=0$),
the angle of perfect absorption $\theta_P$ equals the direction of energy flow. This is not 
necessarily the case just inside BP, where the angle of the energy flow inside BP
at BP/vacuum interface obeys the simple relation:
\begin{align} \label{BPrelation}
\tan\theta_S = \Re\left\{\frac{1}{\varepsilon_{1z}}\right\}\tan \theta_{P}.
\end{align}

Having established the methods for determining the absorption and reflection coefficients, 
we now consider a range of material and geometrical parameters that leads to perfect absorption 
in the low-permittivity regime. To clarify  the coupling of the incident beam to fast wave modes, it is 
important  to examine the corresponding waveguide modes of the structure. The poles of the 
reflection coefficient, where the denominator in  Eq.~(\ref{arrr}) vanishes, yield the allowed  modes:
 \begin{align}
 \tan \left(k_{1z} d \right)+\frac{i\varepsilon_{1z}  k_{0z} k_{1z}} {\varepsilon_{1z} k_0^2 - k_{0x}^2}=0. \label{disp}
\end{align}
The transcendental equation [Eq.~(\ref{disp})] provides four types of solutions 
for the propagation constant $k_{0x}$ due to the $\pm$ signs for $k_{1z}$ and $k_{0z}$ \cite{leaky}. 
The branch leading to perfect absorption corresponds to both $k_{1z}<0$ and $k_{0z}<0$,
yielding fast-wave ($k_{0x}/k_0 < 1$), non-radiative ($k_{0x}\in \Re $) modes that represent a 
coherent superposition of waves that propagate without loss along the BP surface  \cite{feng}. 
Once the propagation constants are found, they can be correlated with the parameters that lead 
to the angles of perfect absorption, $\theta_P$, via $\theta_P=\arcsin(k_{0x}/k_0)$. As an 
alternative approach for finding perfect absorption, we also match the effective field-impedance
of the incident plane wave in free space, ${\cal Z}_0$, to that of BP structure ${\cal Z}_1$, 
where we define  ${\cal Z}_i=E_{xi}/H_{yi}|_{z=0}$. These solutions can then be compared with 
the waveguide modes found in Eq.~(\ref{disp}).

\section{results and discussions}\label{results}
In what follows, we study two scenarios for a strained device. In the first 
case, we assume that an externally applied strain is distributed uniformly throughout the system. 
In the second  case, 
a linearly distributed strain model is implemented to address an 
example of nonuniformly strained devices. 
\subsection{Uniformly strained system}
We have computed the permittivity tensor $\rttensor{\varepsilon}_1(\omega)$ of bulk BP by the 
DFT method. Since BP belongs to the orthorhombic point group $\rm D_{2h}$, there should be,
in principle, differences in the EM response when strain is applied along either the $x$ or $y$ 
directions. For simplicity, we consider here situations where the in-plane strain is applied equally 
in the $x$ and $y$ directions, so that $\epsilon_{xx} = \epsilon_{yy} = \epsilon_\parallel$.
The strain is varied in increments of $2\%$, ranging from $-10\%$ to $+10\%$  for both the 
in-plane $\epsilon_{\parallel}$ and perpendicular strains $\epsilon_{zz}$. Within a simple
Drude model formalism, ENZ responses occur in small regions around the plasma frequency,
and for bulk BP, there are several frequencies around which the permittivity is zero. Thus, 
depending on the strain values, multiple ENZ modes can be accessible over a wide range 
of frequencies. This adds to the fact that the diagonal components of $\rttensor{\varepsilon}_1(\omega)$ 
in general have real parts that vanish at different frequencies. For coherent  perfect absorption 
to take place, all nonzero components of $\rttensor{\varepsilon}_1(\omega)$ take part in the 
EM response, however $\varepsilon_{1z}$  plays the greatest role due in part to the finite-size 
effects (in the $z$-direction) and the corresponding $z$-component of the electric field 
contributing to the necessary interference effects responsible for perfect absorption. 

\begin{figure}[t!] 
\centering
\includegraphics[width=8.5cm,height=5.5cm]{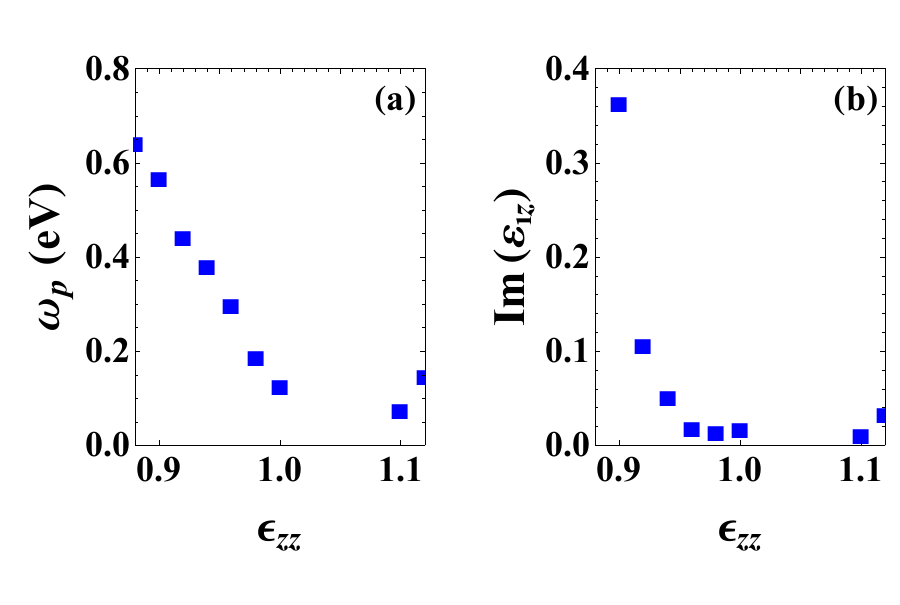}\\
\vspace{-0.8cm}
\includegraphics[width=8.5cm,height=5.5cm]{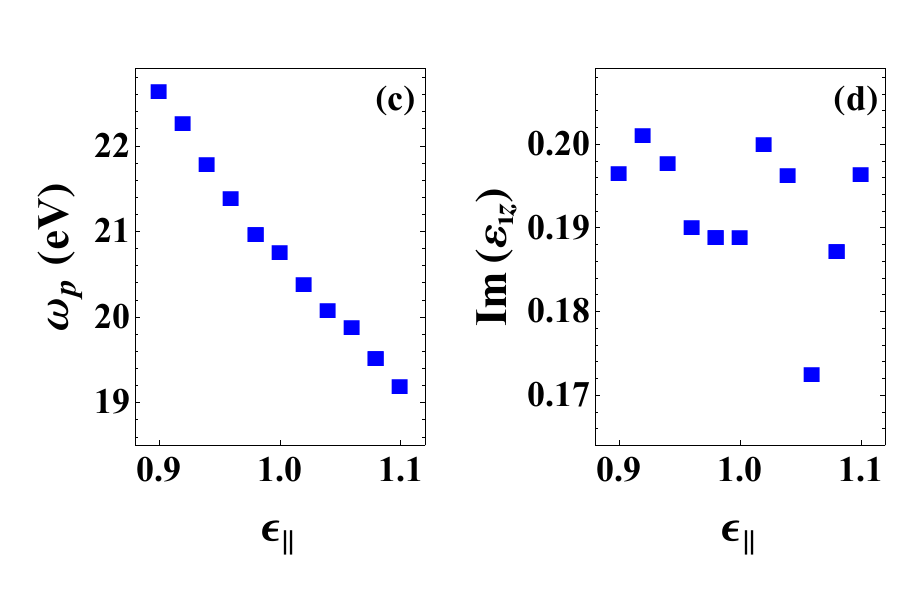}
\caption{(Color online). Top row: (a)-(b) Computed frequencies $\omega_p$
from DFT that lead to an ENZ response in bulk black phosphorus as a function of strain $\epsilon_{zz}$. 
The system is unstrained in the plane of the sample ($\epsilon_{\parallel}=1.0$). 
Bottom row: (c)-(d) The same as a function of strain $\epsilon_{\parallel}$ where the system
is unstrained in the direction normal to the plane of the sample ($\epsilon_{zz}=1.0$)
}
\label{ENZ1}
\end{figure} 

\begin{figure}[t!] 
\centering
\includegraphics[width=8.5cm,height=8.5cm]{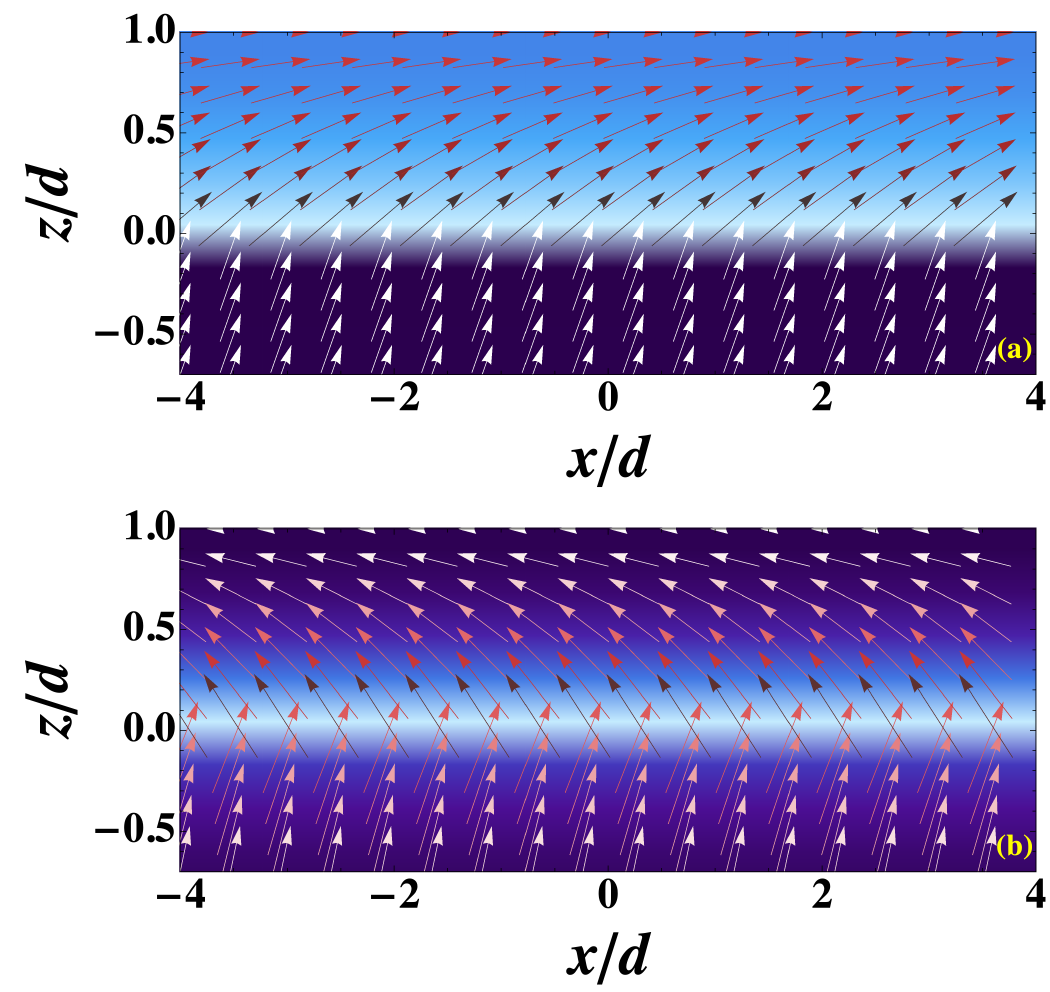}
\caption{
(Color online).  
Directional control of energy flow through strain: Two-dimensional maps illustrating the behavior 
of the time-averaged Poynting vector for a plane wave incident on a black phosphorus film with metallic substrate. The incident angle 
corresponds to $\theta = 34.1^\circ$, and the frequency is set at $20.7 \, {\rm eV}$. In (a) the strain 
parameters correspond to $\epsilon_{\parallel} = 1.04$, leading to perfect absorption, while for (b)
the system is unstrained ($\epsilon_{\parallel}=1$). For both cases there is no strain in the $z$ 
direction ($\epsilon_{zz}=1$). The film thickness is $d=5.42\times10^{-3}\,\upmu {\rm m}$.
The interface separating the black phosphorus from the vacuum region is located at $z=0$ and the vacuum region 
corresponds to $z<0$.
 }
\label{a_vs_theta}
\end{figure} 

\begin{figure}[t!] 
\centering
\includegraphics[width=0.47\textwidth]{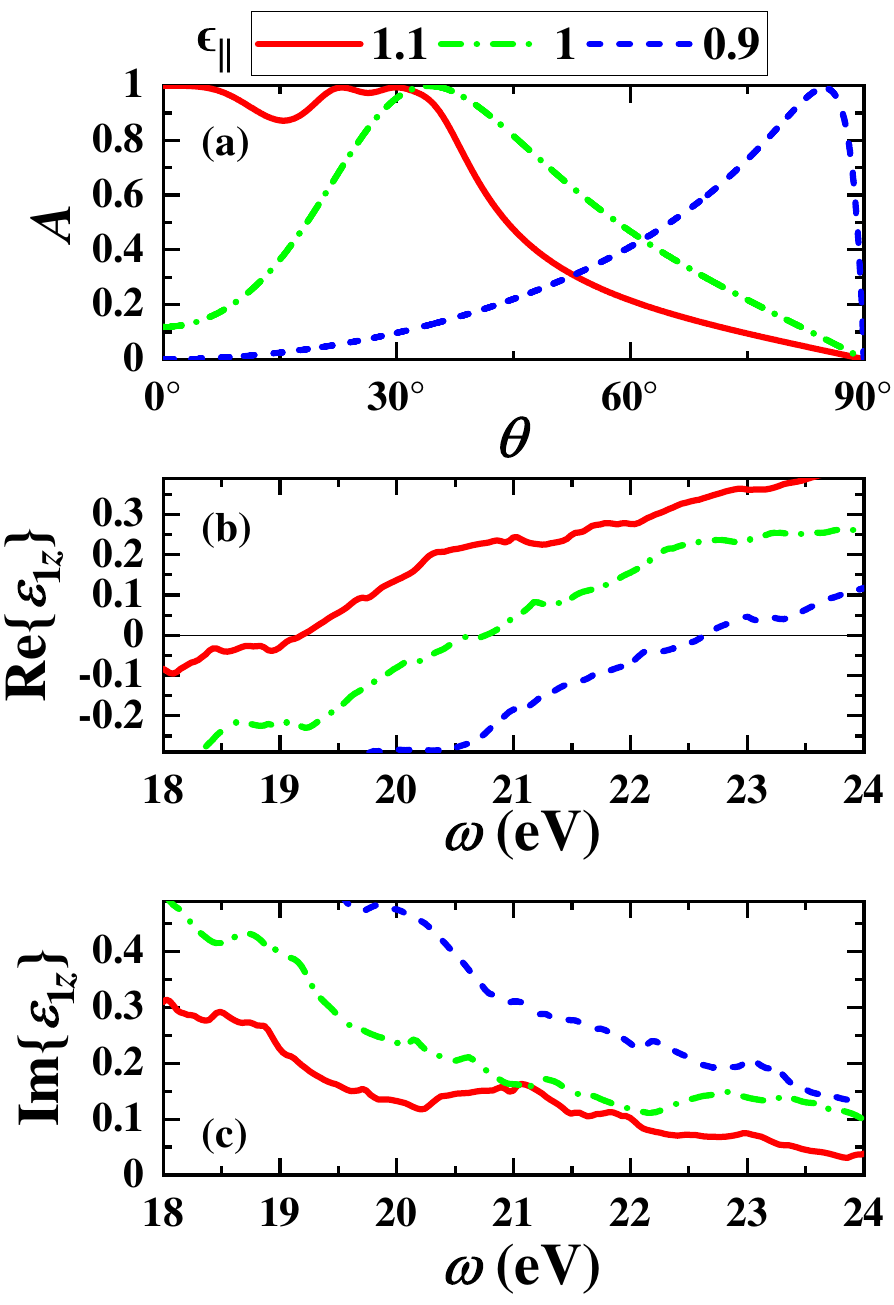}
\caption{(Color online).  
(a) Absorptance as a function of the incident angle $\theta$ for three in-plane strains 
$\epsilon_{\parallel}$, as shown in the legend. The frequency is set at $\omega = 22.6\,{\rm eV}$.
(b, c) The corresponding real and imaginary parts of the permittivity component 
$\varepsilon_{1z}$ (obtained by DFT) are shown as a function of the incident wave frequency.
}
\label{a_vs_theta2}
\end{figure}

To illustrate the  relationship between the ENZ response and strain, we show in
Fig. \ref{ENZ1}(a), the ENZ frequency $\omega_p$ where ${\rm Re}\{\varepsilon_{1z}(\omega_p)\}=0$, 
plotted as a function of strain in the $z$ direction. The corresponding imaginary part of the dielectric response
$\varepsilon_{1z}(\omega)$ is shown in Fig. \ref{ENZ1}(b). Here, we have considered a low 
frequency regime, i.e., $0.1\,{\rm eV}\lesssim\omega\lesssim 0.6\,{\rm eV}$, although ENZ 
scenarios arise for lower frequencies as well. From Fig. \ref{ENZ1}(a), it is seen that $\omega_p$ 
declines nearly linearly as the compressive strain relaxes from $\epsilon_{zz}$=0.9,
until reaching the unstrained state at $\epsilon_{zz}=1$. The imaginary component in 
Fig. \ref{ENZ1}(b) also declines, but much more rapidly. Indeed, the application of a compressive 
strain less than $6\%$ results in a vanishingly small imaginary part of $\varepsilon_{1z}(\omega)$ 
for these ENZ modes, which is a favorable situation for coherent perfect absorption. \cite{feng}
When optimizing BP absorber and limiting undesirable Joule heating, we focus on low to 
moderate dissipation, so that only data with ${\rm Im}(\varepsilon_{1z})\leq 0.4$ are shown. 

In Fig. \ref{ENZ1}(c,d), the system is now unstrained in the $z$-direction, and there are compressive 
and tensile strains within BP plane. A higher frequency regime is also considered where 
$19\,{\rm eV}\lesssim\omega\lesssim 23\,{\rm eV}$. Moving from a compressive to 
tensile strain [Fig. \ref{ENZ1}(c)] causes the ENZ frequency to decline nearly linearly over the whole
range of strains. The imaginary component of the dielectric response $\varepsilon_{1z}$ [Fig. \ref{ENZ1}(d)] reflects slight 
changes in the dissipative nature of the system as the in-plane strain varies. However, the effect is 
weaker at these higher frequencies compared to what was observed in Fig.~\ref{ENZ1}(b).
Nonetheless each scenario considered here reveals that strain has a considerable impact
on the electromagnetic response of our BP absorber. It is worth mentioning that an in-plane tensile 
strain of $10\%$ can open a gap in the band structure of bulk BP as shown in Fig. \ref{bs}(c) of 
Appendix \ref{appendix2}. The band structure plots demonstrate a direct link between the band 
crossings (and gap opening) feature of strain to the dissipation part of the dielectric response. As 
the number of band crossings increases, the probability rate for interband transitions grows, and 
consequently, BP hosts higher loss rates (see Appendix \ref{appendix2}).   
  
To depict how strain can influence energy flow in BP/metallic system, we plot the spatial 
profile of the time-averaged Poynting vector in Fig.~\ref{a_vs_theta}. The arrows indicate the  
direction of energy flow. The BP layer has a thickness $d=5.42\times10^{-3}\,\upmu {\rm m}$,
which has been normalized between $0<z/d<1$, while the vacuum region occupies the space 
$z/d<0$ (see Fig.~\ref{diagramxz}). In Fig.~\ref{a_vs_theta}(a), a $4\%$ in-plane tensile strain 
is applied to the BP plane where the incident wave makes a representative angle of 
$\theta=34.1^\circ$ at the frequency of $\omega=20.7\,{\rm eV}$. The chosen angle corresponds 
to the perfect absorption angle $\theta_P$, and thus all incident EM energy goes into BP layer.
For the given value of the tensile strain, the permittivity tensor of BP has components, 
$\varepsilon_{1z}\approx0.092+0.145 i$, and $\varepsilon_{1x}\approx0.105+0.192 i$.
From Eq.~(\ref{BPrelation}), we find that the incident EM wave undergoes significant refraction 
immediately after entering BP with the wave energy directed at $\theta_S \approx 64^\circ$.
Further inside BP medium, the Poynting vector of the incident EM wave bends increasingly 
until the energy flows nearly parallel ($\theta_S \rightarrow 90^\circ$) close to the surface of the 
metal ($z/d \approx 1$). In Fig.~\ref{a_vs_theta}(b) 
the strain is switched off, producing a dielectric response 
 with $\varepsilon_{1z}\approx-0.085+0.230 i$, and $\varepsilon_{1x}\approx0.073+0.265 i$,
and the rest of the parameters remain intact. The exertion of only $4\%$ of in-plane tensile 
strain changes the energy flow direction by nearly $180^\circ$. This follows from the 
strain-induced sign change of the real part of  $\varepsilon_{1z}$. Near the interface in the 
vacuum, we find from Eq.~(\ref{vrel}) that the direction of the net flow of energy is shifted slightly to 
$\theta_S \approx 38.5^\circ$ as BP now partially reflects some of the energy of the incident wave.

The BP perfect absorber in Fig.~\ref{diagramxz}  can be tailored to absorb EM energy over a wide
range of incident angles. To illustrate this, Fig.~\ref{a_vs_theta2}(a) shows the absorptance of the 
incoming plane wave as a function of its incident angle $\theta$.  Each curve represents a different 
in-plane strain value of the two-dimensional BP sheet (see legend). The strains 
$\epsilon_{\parallel}=0.9, 1.0$, and $\epsilon_{\parallel}=1.1$ correspond to BP film thicknesses
$d=1.5\times10^{-4}\upmu{\rm m}$, $d=0.012\,\upmu{\rm m}$, and $d=0.18\,\upmu{\rm m}$, 
respectively. We will show below that the thickness $d$ and angle $\theta$ have an intricate 
relationship that must be satisfied to achieve perfect absorption.
 
\begin{figure*}[t!] 
\centering
\includegraphics[width=0.96\textwidth]{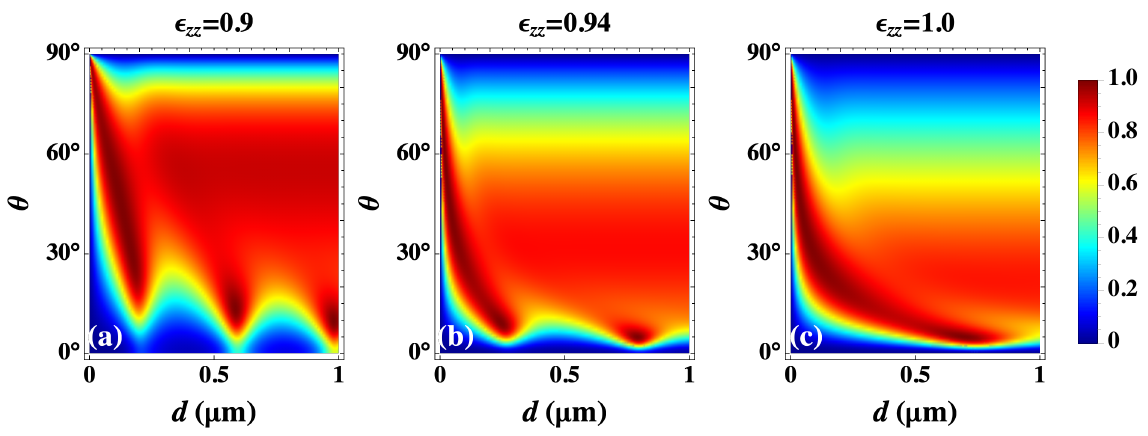}\\
\includegraphics[width=0.96\textwidth]{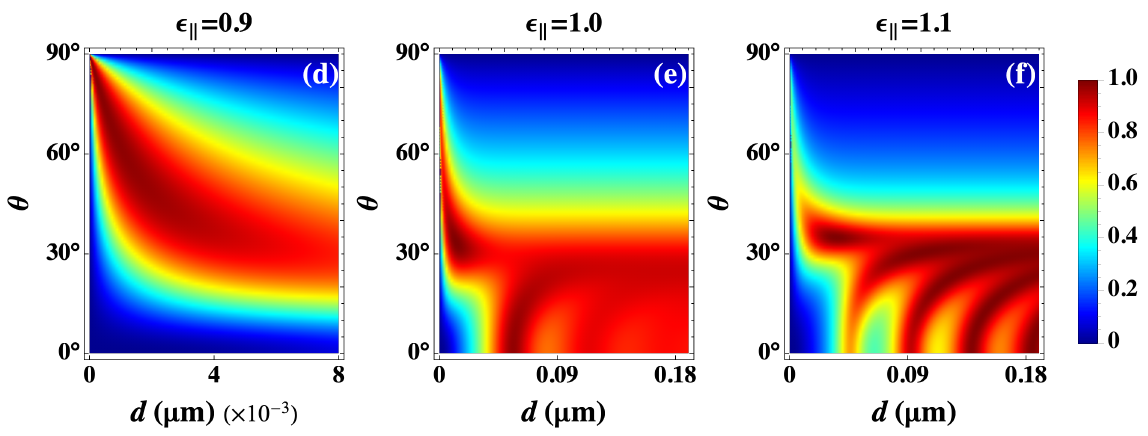}
\caption{(Color online).  
Color maps representing the absorptance as a function of the angle of incidence and 
thickness of a black phosphorus film with metallic substrate}. Six different strains are considered as labeled. 
Top row: The parameters used correspond to the strains and frequencies calculated in 
Fig.~\ref{ENZ1}(a,b) where there is only out-of-plane strain ($\epsilon_{\parallel}=1$). 
Bottom row: Only in-plane strain is present ($\epsilon_{zz}=1$) and the frequency is set 
to $\omega = 22.6\,{\rm eV}$.
\label{theta_vs_d_dyy1}
\end{figure*}       
To understand the angular dependence of the absorption, it is important to have an accurate 
characterization of the EM response in BP, in particular $\varepsilon_{1z}$. For perfect 
absorbers in the ENZ or low permittivity regimes, the imaginary part of the permittivity plays 
a crucial role in determining which incident angles are the perfect absorption angles.
Indeed, while optimizing system parameters for practical devices, it is important to keep in 
mind that when there is a low dissipation in the medium, a very thin layer is required to achieve 
perfect absorption.\cite{} As the dissipation increases (e.g., due to changing $\omega$ or 
$\epsilon_{zz}$ and $\epsilon_{\parallel}$) thicker BP layers are needed. Thus, it is imperative to 
clarify the frequency dispersive nature of permittivity. In Fig.~\ref{a_vs_theta2}(b,c), the real and 
imaginary parts of $\varepsilon_{1z}$ are shown as a 
function of frequency. When no strain is exerted [Fig. \ref{a_vs_theta2}(a)], the
maximum absorption occurs in the range $35^\circ\lesssim\theta\lesssim45^\circ$. 
A tensile strain of $10\%$ then widens
 the range of  perfect absorption angles to fall within
$0^\circ\lesssim\theta\lesssim40^\circ$. 
Upon changing the type of strain to compressive, 
near perfect absorption  arises for incident waves at near grazing, 
with  $\theta\approx85^\circ$.
These results demonstrate that the application of 
a low strain along the principal crystallographic directions of BP structure  can effectively 
control the perfect absorption of the incident beam. This occurs for a specific incident angle 
that correlates with the proper ENZ modes (see Fig.~\ref{ENZ1}). 

\begingroup
\squeezetable
\begin{table}[h] 
\begin{ruledtabular}
\caption{\label{epstab}
Dielectric  response and strain parameters for bulk BP calculated by DFT. The strains 
($\epsilon_\parallel$, $\epsilon_{zz}$) and permittivity components 
($\varepsilon_{1x}$, $\varepsilon_{1z}$) are presented for the given frequency ($\omega$).
}
\centering 
\begin{tabular}{c c c c c} \\ 
 $\omega$(eV) & $\epsilon_\parallel$ & $\epsilon_{zz}$ &$\varepsilon_{1z}$ & $\varepsilon_{1x}$
 \\ [1ex]
 \hline \vspace{-.1cm}  \\ 
0.573 & 1.00 &0.90& $0.000996+0.374 i$  &$7.577 + 0.152 i$   \\
0.380 & 1.00 & 0.94 & $0.000919+0.0527 i$  &$9.419 + 0.0371 i$  \\
0.124 & 1.00 & 1.00 & $0.00153+0.0176 i$ & $11.294+0.00779 i  $ \\
22.64 & 0.90 & 1.00 & $0.000149+0.196 i$ & $0.160+0.140 i $ \\
22.64 & 1.00 & 1.00 & $0.233 + 0.142 i$ & $0.198 + 0.0979 i$  \\
22.64 & 1.10 & 1.00 & $0.341 + 0.0688 i$ & $0.347 + 0.0402 i$ \\
\end{tabular}
\end{ruledtabular}
\end{table}
\endgroup

To further shed light on this phenomenon, 
in  Fig.~\ref{theta_vs_d_dyy1} we plot the absorptance as a function of the incident beam 
angle and BP thickness. The  strain values are labeled in each figure, and the top row 
of Fig.~\ref{theta_vs_d_dyy1} displays the strain applied orthogonal to the planar interface 
(along the $z$-direction), whereas in the bottom row in-plane strain is applied 
(in the $\var{x-y}$ plane). The relevant components of the permittivity tensor for the 
considered strains and frequencies are listed in Table~\ref{epstab}. For our geometry and 
polarization state of the incident wave, the component $\varepsilon_{1x}$ plays a limited 
role in the absorption of EM energy, but it is listed for completeness.

For the top row in Fig.~\ref{theta_vs_d_dyy1}, we consider the compressive strains
(a) $\epsilon_{zz} = 0.9$ and (b) $\epsilon_{zz}=0.94$, while (c) shows the unstrained 
case with $\epsilon_{zz}=1$. The frequencies have been chosen to correlate with 
$\omega_p$ in Fig.~\ref{ENZ1}(a), so that an ENZ response is generated for each strain 
value. Thus, we have used the values (a) $\omega=0.57$, (b) $\omega=0.38$, and (c) 
$\omega=0.12\,{\rm eV}$, with each generating different  levels of dissipation according to  
Fig.~\ref{ENZ1}(b) (see also Table~\ref{epstab}). Since  the dissipative response declines 
as the strain parameter changes  from an unstrained state,  $\epsilon_{zz}=1.0$,
to  compressive at $\epsilon_{zz}=0.9$, we have a controllable platform in which EM  
absorption is dependent on both frequency and strain. 

In Fig. \ref{theta_vs_d_dyy1}(d-f),  the frequency of the incident wave is now fixed at 
$\omega = 22.6\,{\rm eV}$ while the system goes from compressive strain with 
$\epsilon_{\parallel}=0.9$  to tensile strain with $\epsilon_{\parallel}=1.1$  For the 
tensile strain case, there is a small real part in the permittivity component 
$\varepsilon_{1z}$, which vanishes as the system undergoes compressive strain. This ENZ state generates a moderate amount of dissipation which weakens as the 
in-plane strain parameter $\epsilon_\parallel$ increases. 

The results in Fig. \ref{theta_vs_d_dyy1} reveal that both compressive and tensile 
strain can generate perfect absorption over a wide range of $\theta$ and thickness 
values. In particular, Figs. \ref{theta_vs_d_dyy1}(a-c) shows that as $\epsilon_{zz}$
increases, there exists a broader range of incident angles that result in perfect 
absorption. This follows from a reduction in the dissipative response of $\varepsilon_{1z}$
as the compressive strain is reduced (see Fig.~\ref{ENZ1}). This behavior is consistent 
with Weyl semimetal absorbers that have tunable dissipation\cite{WS,nws1,nws2,nws3,nws4,nws5,nws6,nws7} and anisotropic 
ENZ coherent perfect absorbers.\cite{feng} Overall, for an incoming wave at near grazing 
incidence ($\theta \sim 90^\circ$) to be fully absorbed, very thin subwavelength BP layers 
are needed, while for near normal incidence ($\theta \sim 0^\circ$), thicker layers are 
required.

We turn now back to the cases with lateral strain and fixed frequency  $\omega = 22.6\,{\rm eV}$.
As Fig.~\ref{theta_vs_d_dyy1}(d) reveals, when there is a compressive in-plane strain, 
the high absorption region appears for thin BP layers and a range of incident angles 
satisfying  $\theta \gtrsim 50^\circ$. As the strain parameter $\epsilon_\parallel$ increases 
[Fig.~\ref{theta_vs_d_dyy1}(e,f)], the perfect absorption regions become limited to smaller 
angles of incidence and larger BP layer thicknesses. Correspondingly, there is a substantial 
increase in the EM modes responsible for complete absorption of the incident wave for tensile 
strain [Fig.~\ref{theta_vs_d_dyy1}(f)], which results in a greater range of permitted BP thicknesses.

The compressive and tensile strains (of magnitude less than $\lesssim 4\%$)
  considered throughout the paper present experimentally accessible regimes for fabricating a
  perfect-absorber device. We have found that a thin spacer layer in between the
  BP and metallic substrate often has
  little effect on the results. Therefore, one possible configuration 
  for controlling the in-plane strain of the BP film could involve 
  an elastomeric
  spacer or matrix 
  containing 
  BP stack and metal. 
With proper tailoring of the elastomeric vertical edges, 
    equal strain can be exerted throughout BP layer.
  For larger strains, we assume that 
  the bulk BP undergoes strain values that do not exceed $10\%$,
  thus maintaining the more energetically stable allotrope, consistent with previous theoretical
  studies. \cite{strain1, strain2, strain3} 
  Nevertheless, exerting strain values on the order of
  $10\%$ can be challenging in practice with 
   current experimental capabilities\cite{strain3}. For larger systems, 
   the strain may become inhomogeneous under specific situations. In this case, experimental guidance for
   modeling the explicit spatially inhomogeneous strain pattern seems necessary as there are numerous options 
   to consider. Moreover, this problem goes beyond first-principles calculations as millions of atoms might be involved and other approaches such as empirical potential or effective Hamiltonian treatments should be 
   employed for calculating the dielectric response of such
   systems. In the next section, 
   we expand the multiscale approach presented above to spatially nonuniform strain patterns.   
 
 \begin{figure}[t!] 
\centering
\includegraphics[width=0.47\textwidth]{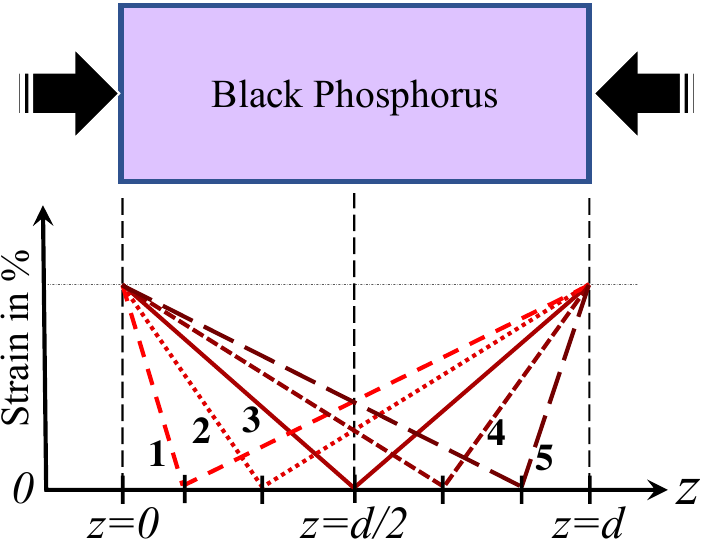}
\caption{(Color online). Inhomogeneous strain in the black phosphorous is 
accounted for by using  a linear $z$-dependent model. The spatially inhomogeneous strain develops along the $z$ direction 
and it varies linearly (red lines)
 along the strain direction from a maximum (horizontal dashed line) at locations
  $z=0$  and $z=d$, to zero strain at points marked by the vertical small indicators at $z=d/6, d/3, d/2, 2d/3, 5d/6$. The pertaining curves to these five linear models are marked by $1,2,3,4,5$.
  The arrows portray  the direction of applied strain
  that propagates linearly inside the BP region confined between $z=0$ and $z=d$.}
\label{diagram2}
\end{figure} 

 \begin{figure*}
\centering
\includegraphics[width=0.85\textwidth]{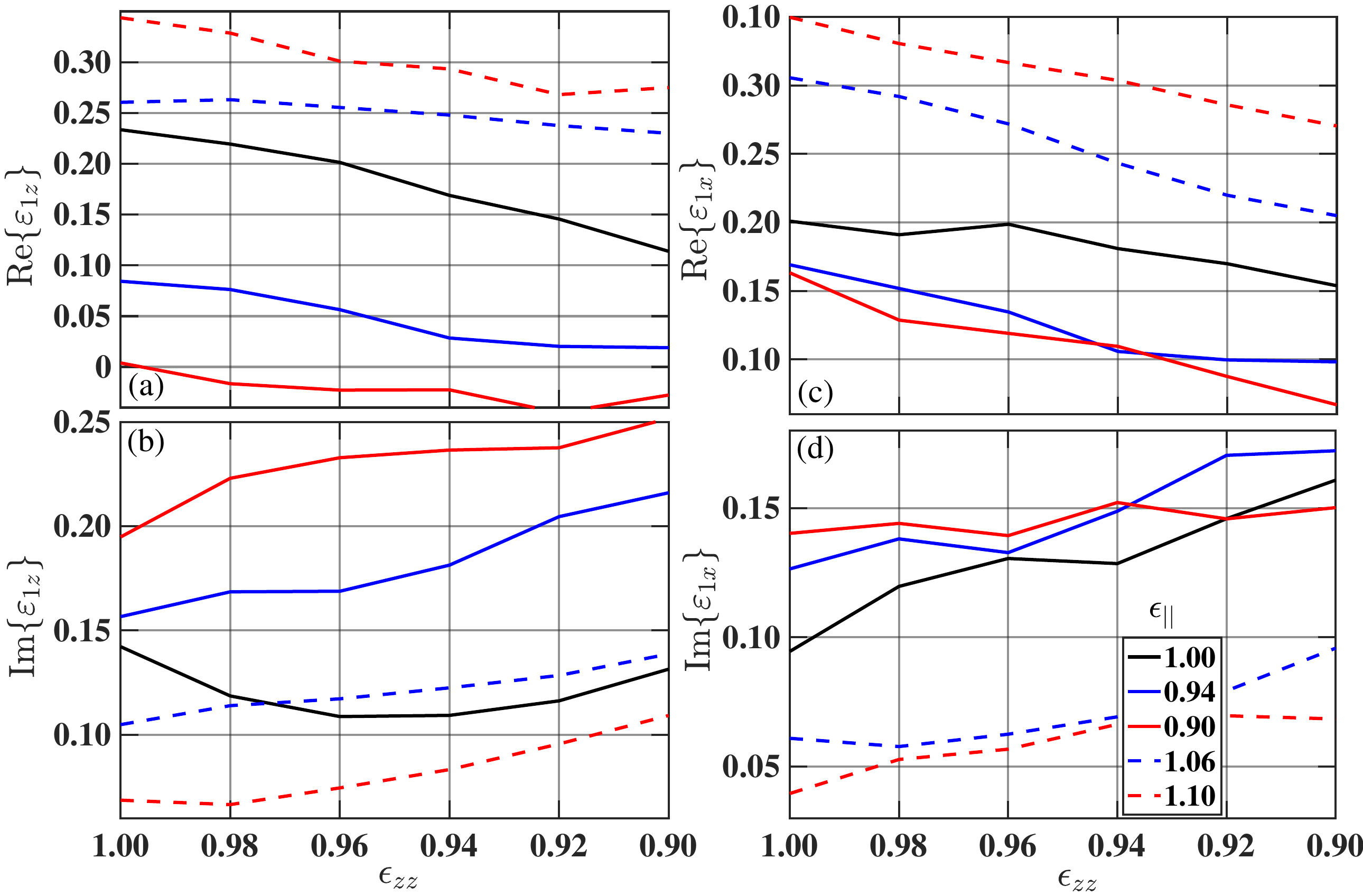}
\caption{(Color online). The permittivity of BP as a function of compressive strain $\epsilon_{zz}$ from $0$ to $-10\%$. The panels (a) and (b) show the real and imaginary parts of $\varepsilon_{1z}$ whereas the panels (c) and (d) are the real and imaginary part of $\varepsilon_{1x}$. The in-plane strain spans the
strain-free, compressive, and tensile strain regimes; $\epsilon_{||}=1.0, 0.94, 0.90, 1.06, 1.10$. }
\label{e1xz_zstrain}
\end{figure*}

\subsection{Inhomogeneously strained system} \label{inhomogeneous}

As is seen above, by modifying the permittivity tensor, an externally applied strain can considerably change the response of a material to an incident electromagnetic wave. If the applied strain within the 
material is spatially distributed in a nonuniform way, the permittivity tensor 
becomes spatially inhomogeneous as well. To simulate 
an inhomogeneously strained system, 
one should resort to Maxwell's equations with  location-dependent permittivity $\rttensor{\boldsymbol{\varepsilon}}_n(\bf{r})$ and permeability $\rttensor{\boldsymbol{\mu}}_n(\bf{r})$:
\begin{equation}
\label{mxwl_inhomo} 
\begin{aligned} 
&\boldsymbol{\nabla}\times\left(\rttensor{\boldsymbol{\mu}}_n^{-1}(\bf{r})\boldsymbol{\nabla}\times \bm{E}(\bf{r})\right) = \omega^2\mu_0\varepsilon_0\rttensor{\boldsymbol{\varepsilon}}_n(\bf{r})\bm{E}(\bf{r}),\\
&\boldsymbol{\nabla}\times\left(\rttensor{\boldsymbol{\varepsilon}}_n^{-1}(\bf{r})\boldsymbol{\nabla}\times \bm{H}(\bf{r})\right) = \omega^2\mu_0\varepsilon_0\rttensor{\boldsymbol{\mu}}_n(\bf{r})\bm{H}(\bf{r}).
\end{aligned}
\end{equation}
The above equations describe 
the behavior of the electromagnetic fields $\bf E$, $\bf H$ for
a generic system with a spatially-inhomogeneous 
electromagnetic response. 
To be able to make use of Eqs.~(\ref{mxwl_inhomo}),
we consider simple linear models for the spatial behavior of the permittivity,
as shown in
 Fig. \ref{diagram2}, and the models are marked by $1,2,3,4,5$.
 The
strain takes its maximum value at the interfaces
surrounding BP ($z=0,d$), 
and then linearly declines to zero at $z=d/6, d/3, d/2, 2d/3, 5d/6$ corresponding to models $1,2,3,4,5$, respectively.
By considering
the  model described above  for an inhomogeneous strained system
and the
 fact that BP is a nonmagnetic material,
the  permeability and permittivity tensors reduce to 
$\rttensor{\boldsymbol{\mu}}_n(\bf{r})=\rttensor{\boldsymbol{1}}$ and 
$\rttensor{\boldsymbol{\varepsilon}}_n({\bf r})=\rttensor{\boldsymbol{\varepsilon}}_n({z})$. 
We consider the same
form for the  incident electromagnetic field, given by Eqs.~(\ref{EHfields}),
 in the inhomogeneously strained system 
 as that considered in the previous section for uniformly strained BP.
 By incorporating the above assumptions
 into Eq.~(\ref{mxwl_inhomo}), we arrive at the following 
 differential equation for the $H_y$ field:
\begin{align}\label{pdeHy}
&\frac{d^2H_y(z)}{dz^2} + \varepsilon_{1x}(z)\frac{d \varepsilon_{1x}^{-1}(z)}{dz}\frac{dH_y(z)}{dz} + \nonumber\\&\left(\omega^2\mu_0\varepsilon_0\varepsilon_{1x}(z) - \frac{\varepsilon_{1x}(z)}{\varepsilon_{1z}(z)}k_x^2\right)H_y(z)=0.
\end{align}  
After solving Eq.~(\ref{pdeHy}) with the boundary conditions given in Sec.~\ref{method}, the $E_x(z), E_y(z)$ components of the electric field can be obtained by substituting the $H_y(z)$ field into the original Maxwell's equations, i.e., Eqs.~(\ref{mxwl_homo}). 

\begin{figure*}[t!] 
\centering
\includegraphics[width=1.0\textwidth]{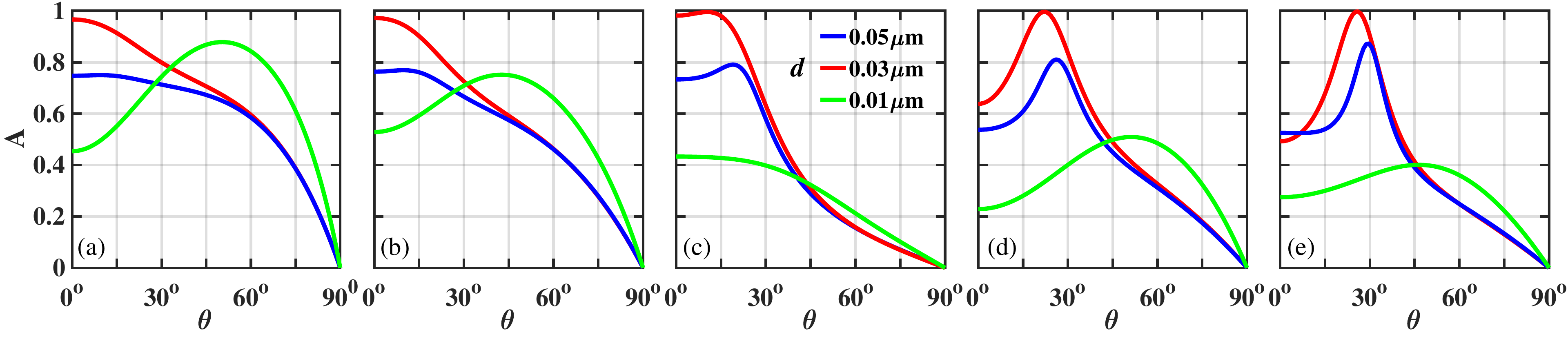}\\
\includegraphics[width=1.0\textwidth]{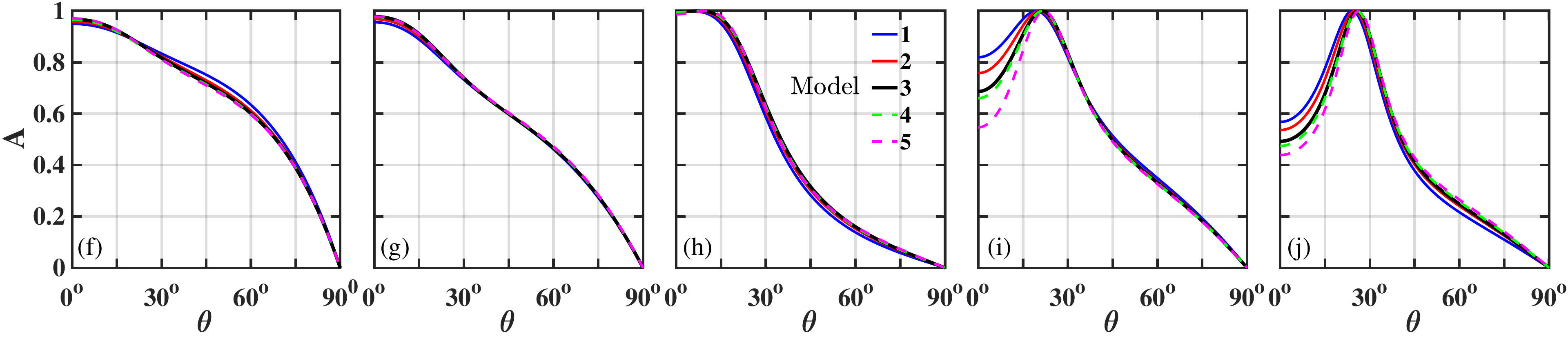}
\caption{(Color online). Absorptance vs the incident angle $\theta$ of electromagnetic wave. The frequency of the electromagnetic wave is set to 
$\omega=22.6$~eV. The in-plane strain increases from the leftmost column to the rightmost column, i.e., $\epsilon_{||}=-10\%, -6.0\%, 0.0\%, +6.0\%, +10\%$ in (a,f), (b,g), (c,h), (d,i), and (e,j), respectively. In the top row panels, three different thicknesses of BP are set; $\rm {\it d}=0.5\mu m, 0.3\mu m, 0.1\mu m$, when model 1 is implemented. In the bottom row panels, the thickness of BP is set to $\rm {\it d}=0.3\mu$ and the five different models (marked by $1,2,3,4,5$) are implemented.}
\label{A_inhomo}
\end{figure*}

For a sufficiently large
 BP sample, the local permittivity at a certain location with a certain strain (according to the above linear model) can be approximately calculated with DFT
 using 
  the bulk black phosphorus unit cell. 
  Following this approach yields  the spatial profile for the permittivity, $\varepsilon_{1z}(z)$ 
  shown in Fig.~\ref{e1xz_zstrain}. 
  To proceed towards  tractable solutions,
  we  assume that the applied strain in the $x$ and $y$ directions is
  uniform so that the parallel wavevector component  $k_x$ remains a conserved quantity upon the scattering process. 
  Note that otherwise, Maxwell's equations (\ref{mxwl_inhomo}) result in three-dimensional position-dependent partial differential equations, making any subsequent analysis highly complicated.
  
 Incorporating the linear  model  for $\varepsilon_{1z}(z)$ into Eq.~(\ref{pdeHy}), we study the absorption properties of this system in Fig.~\ref{A_inhomo}. The absorptance $A$ is plotted against the  incident angle 
 $\theta$, when the in-plane strain (strain exerted in the $xy$ plane) can have a finite value. In Figs. \ref{A_inhomo}(a,f), \ref{A_inhomo}(b,g), \ref{A_inhomo}(c,h), \ref{A_inhomo}(d,i), and \ref{A_inhomo}(e,j), the in-plane strain is set to $- 10\%$ ($\epsilon_{\parallel}=0.90$), $-6 \%$ ($\epsilon_{\parallel}=0.94$), $0\%$ ($\epsilon_{\parallel}=1.0$), $-+6\%$ ($\epsilon_{\parallel}=1.06$), and $+ 10\%$ ($\epsilon_{\parallel}=1.10$), respectively. The maximum of strain in the $z$ direction at the boundaries $z=0,d/2$ are set to a representative value, i.e., $-10\%$ that linearly declines toward the middle of the BP layer according to the models described in Fig. \ref{diagram2}. The top row panels in Fig. \ref{A_inhomo} illustrate the absorptance when
 BP possesses a thickness of $d=0.5\mu$m, $0.3\mu$m, and $0.1\mu$m when model 1 is implemented. 
 As is clearly seen, the perfect absorption found in the uniform case
  [Fig.~\ref{a_vs_theta2}(a)]
   can also be achieved by manipulating the thickness of the BP layer $d $ in the inhomogeneous model scenario considered. As in the homogenous strain case, the in-plane strain can control the angle of perfect absorption although now limited to $\theta\lesssim 30^\circ$. Also, the results reveal that for the linearly inhomogeneous strain model implemented, the system absorbs the incident electromagnetic wave the most when the thickness of the BP layer is around $d=0.3\mu$m. In the bottom row panels of Fig. \ref{A_inhomo}, the thickness of the BP layer is set fix to $d=0.3\mu$m and the five different strain models shown in Fig. \ref{diagram2} are implemented. The results show that the different inhomogeneous strain models keep the perfect absorption of the device almost intact. 
   Our further study (not shown) demonstrates that the multiple perfect absorption peaks found for the $\epsilon_\parallel = 1.1$ cases 
   of tensile strain in 
   Fig.~\ref{a_vs_theta2}(a) reduces 
   as the
   modified effective dielectric response increases the BP reflectivity 
   due to
   the incident wave not fully coupling to the structure. Therefore, by tailoring the thickness of the BP appropriately, perfect absorption can also
   be achieved when there is a  spatially inhomogeneous strain-dependent dielectric response.

\section{conclusion} \label{conclusions} 
We have studied the absorption of electromagnetic energy for a semi-infinite strained bulk black 
phosphorus layer that is deposited on a metallic substrate. Using the density functional theory of 
electronic structure, we obtained the dielectric response tensor of black phosphorus subject to 
compressive and tensile strains along principal crystallography directions. The permittivity along 
the direction normal to the black phosphorus layer was found to exhibit multiple epsilon-near-zero 
conditions in a large frequency range by applying appropriate strain. Incorporating the calculated 
permittivity  tensor, we solved Maxwell's equations for the electromagnetic modes, demonstrating 
that the exertion of strain can switch the direction of electromagnetic wave energy flow within the 
black phosphorus layer. The applied strain was demonstrated as an effective control knob for 
tuning the optical and electronic properties of black phosphorus, resulting in efficient control of 
the absorption of an incident electromagnetic wave with a largely tunable angle. Considering  spatially nonuniform strain
profiles along the direction normal to the black phosphorus layer, we showed
 that the application of strain  can produce near perfect absorption of the incident wave. The presented 
results open up new avenues for the practical use of coherent and perfect absorption over a wide 
range of incident angles, frequencies, and layer thicknesses. 

\acknowledgments 
The DFT calculations were performed using the resources provided by UNINETT Sigma2 - the National Infrastructure for High Performance Computing and Data Storage in Norway. Part of the calculations were performed using HPC resources from the DOD High Performance Computing Modernization Program (HPCMP). K.H. is supported in part by the NAWCWD In Laboratory Independent Research (ILIR) program and a grant of HPC resources from the DOD HPCMP.  

\appendix

\begin{figure}[t!] 
\centering
\includegraphics[width=0.45\textwidth]{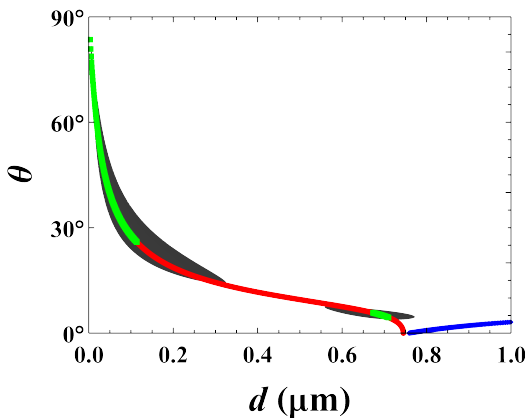}
\caption{(Color online).  
Electromagnetic  modes for  the system in Fig.~\ref{diagramxz} with the 
same parameter values as used in Fig.~\ref{theta_vs_d_dyy1}(c). The red curve depicts the 
fast-wave ($k_{0x}/k_0<1$) waveguide modes, while the green curves correspond to
the impedance matched modes (overlapping with the red curve).
As the thickness increases, eventually radiative leaky-waves emerge 
(blue curve). The angle $\theta$ is determined from the propagation constants 
$k_{0x}$ via $\theta = \arcsin (k_{0x}/k_0)$. For reference, the black region 
is extracted from the high absorptance ($>95\%$) data of Fig.~\ref{theta_vs_d_dyy1}(c).
}
\label{theta_vs_d_CPA}
\end{figure} 

\begin{figure*}[t] 
\centering
\includegraphics[width=18.0cm,height=8.50cm]{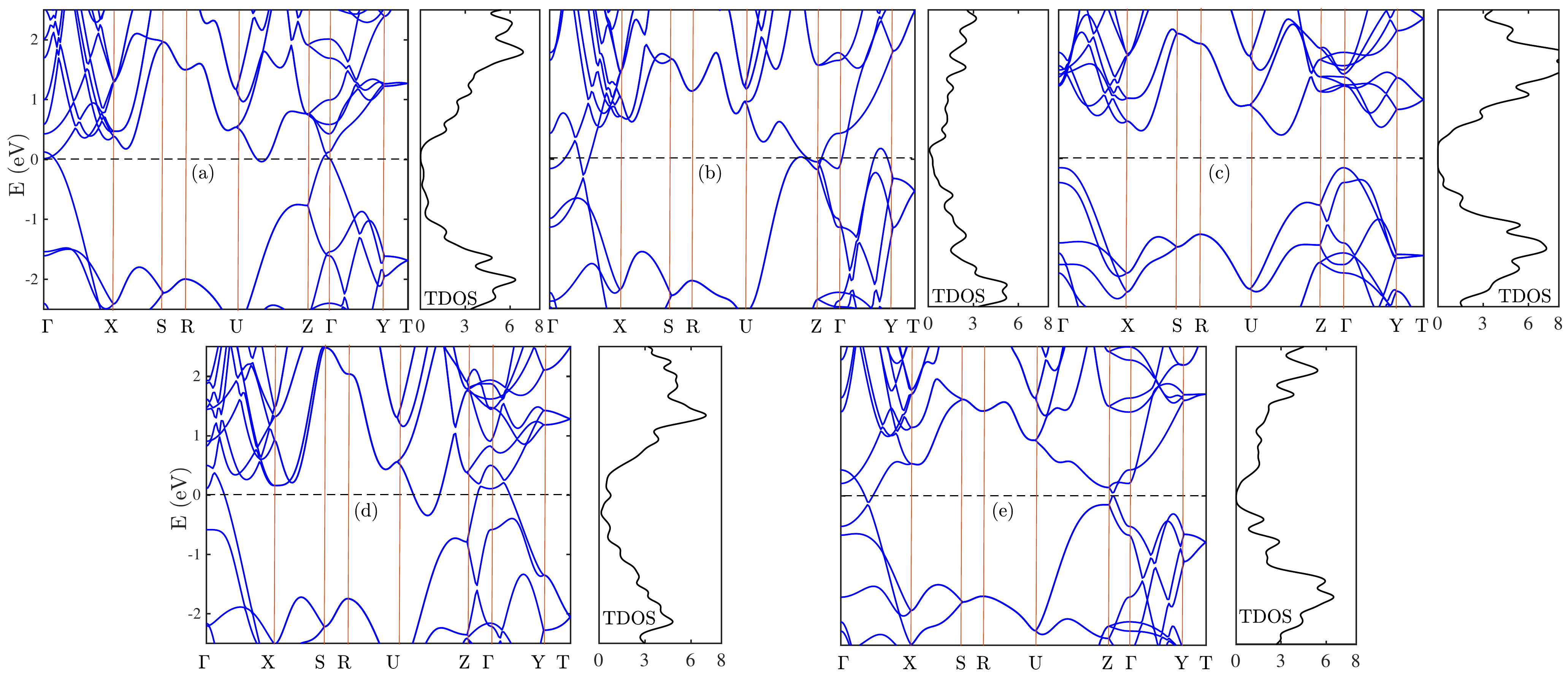}
\caption{(Color online). Electronic band structures and total densities of states 
for bulk black phosphorus. The strain parameters are as follows: 
(a) $\epsilon_\parallel=1.0$, (b) $\epsilon_\parallel = 0.9$,
and (c) $\epsilon_\parallel = 1.1$ for in-plane strain, and
(d) $\epsilon_{zz}=0.9$ and (e) $\epsilon_{zz}=1.1$ for out-of-plane strain.
In (a)-(c) the system is unstrained along $z$ ($\epsilon_{zz} =1$), and
there is no in-plane strain in (d) and (e) ($\epsilon_\parallel=1.0$).
The Fermi energy is located at the energy $E=0$. The axes labels of the top row panels are the same as for the bottom row panels.
}
\label{bs}
\end{figure*} 

\section{Field impedance-matching and waveguide mode approaches for coherent perfect absorption} \label{appendix1}

To gain further insight into the absorption mechanism occurring in the 
BP structure, we consider alternative approaches that can reveal the 
underlying EM modes responsible for the complete absorption of incident 
EM energy. In the case of coherent perfect absorption, the metallic layer 
must generate the proper reflective waves that interfere destructively in 
the BP region. To isolate the absorption effects that arise from purely 
dissipative processes, we have considered in this paper moderate to 
extremely weak amounts of loss. As was previously shown, when the 
structure exhibits an ENZ response and the component of the permittivity 
tensor perpendicular to BP layer exhibits low loss, relatively thin layers 
are required to achieve complete absorption.\cite{feng} To discuss the 
previous findings in Fig.~\ref{theta_vs_d_dyy1} within the context of 
coherent perfect absorption, we implement two independent and 
complimentary methods.

The first method involves matching the field impedance of the incident plane wave
to that of the BP/metal structure. This is achieved by setting ${\cal Z}_0={\cal Z}_1$ at 
the vacuum/BP interface, where the field impedance is defined as
\begin{align}
{\cal Z}_i=\frac{E_{xi}}{H_{yi}} \Biggl |_{z=0}\hspace{-0.15cm},
\end{align}
for either the vacuum or BP regions (identified by $i=0,1$). For the BP/metal system, 
we find
\begin{align}
{\cal Z}_1 = -i \eta_0 \frac{(\epsilon_{1z} k_0^2-k_{0x}^2)\tan(k_{1z} d)}{\epsilon_{1z} k_0 k_{1z}},
\end{align}
whereas for the vacuum region the result is simply ${\cal Z}_0 = \eta_0 k_{0z}/k_0$.
After impedance matching, the resultant expression constrains the allowed geometrical 
and material parameters leading to the reflection coefficient $r$ vanishing [see Eq.~(\ref{arrr})].

An alternative  approach views the  structure in Fig.~\ref{diagramxz} as a waveguide, 
so that the incident plane wave is absent and the electric and magnetic fields in the 
vacuum region decay exponentially. For example, the magnetic field would be written 
$H_{y0} \sim e^{-k_{0z} z} e^{i k_{0x} x}$. As before, the form for the EM fields in the 
BP region are linear combination of waves with the wavevectors given in Eq.~(\ref{kz1}).
After invoking the usual interface and boundary conditions, the inherent guided wave 
modes of the structure can be found. The result is Eq.~(\ref{disp}), which is equivalent to 
finding where the denominator of the reflection coefficient in Eq.~(\ref{arrr}) vanishes.
Since we are interested in perfect coupling of the incident plane wave to the waveguide,
we focus on the fast-wave non-radiative solutions whereby $k_{1z}/k_0<1$. 

The results of the two approaches are presented in Fig.~\ref{theta_vs_d_CPA}, where  
the calculated  perfect absorption modes of BP layered structure are plotted.
All the parameter values used are identical to those of Fig.~\ref{theta_vs_d_dyy1}(c). 
The invariant wavevector component $k_{0x}$ is varied along with $d$,
and each data point is an allowed root to the corresponding transcendental equations. 
The green curves are the calculated  waveguide modes (Eq.~(\ref{disp})), using a 
root-finding algorithm, and the red curves arise from the field impedance matching 
method. As the thickness increases, leaky wave modes arise where the solutions to 
Eq.~(\ref{disp}) admit propagation constants with a finite imaginary component $\alpha>0$.
As can be seen, the results of the two methods are in excellent agreement.    

\begingroup
\squeezetable
\begin{table}[b] 
\begin{ruledtabular}
\caption{\label{abctab}
Lattice parameters ($a_0, b_0, c_0, \alpha, \beta, \gamma$) and the normalized location of phosphorus atoms, $\text{P}_i , \; i=1,2, ..., 8$, ($x,y,z$) for the relaxed bulk black phosphorus unit cell at zero strain.
}
\centering 
\begin{tabular}{c c c c c}  
 $a_0$ ($\textup{\AA}$) & $b_0$ ($\textup{\AA}$) & $c_0$ ($\textup{\AA}$) & Vol ($\textup{\AA}^3$) \\ \vspace{0.1cm}
 3.31590 & 4.50640 & 10.44520 & 156.080247\\
 $\alpha$ (\text{deg}) & $\beta$ (\text{deg}) & $\gamma$ (\text{deg}) & ~ \\
  $90^{\circ}$  & $90^{\circ}$  & $90^{\circ}$  & ~ \\ 
 \hline
 \vspace{0.1cm} Atom & x~($\textup{\AA}$) & y~($\textup{\AA}$) & z~($\textup{\AA}$) \\ 
 $\rm P_1$ & 0.25000 & 0.08486 & 0.07454\\
 $\rm P_2$ & 0.75000 & 0.91514 & 0.37093\\
 $\rm P_3$ & 0.25000 & 0.58486 & 0.37093\\
 $\rm P_4$ & 0.75000 & 0.41514 & 0.07454\\
 $\rm P_5$ & 0.75000 & 0.08486 & 0.57454\\
 $\rm P_6$ & 0.25000 & 0.91514 & 0.87093\\
 $\rm P_7$ & 0.75000 & 0.58486 & 0.87093\\
 $\rm P_8$ & 0.25000 & 0.41514& 0.57454\\
\end{tabular}
\end{ruledtabular}
\end{table}
\endgroup

\begingroup
\squeezetable
\begin{table}[b] 
\begin{ruledtabular}
\caption{\label{abctab2}
Lattice parameters ($a_0, b_0, c_0, \alpha, \beta, \gamma$) for black phosphorus 
undergoing  in-plane and out-of-plane compressive and tensile strains.
\centering 
\begin{tabular}{l c c c c c}  
 ~ & $a_0$ ($\textup{\AA}$) & $b_0$ ($\textup{\AA}$) & $c_0$ ($\textup{\AA}$) & Vol ($\textup{\AA}^3$) \\ 
 \underline{In-plane}\\
 \text{10$\%$ Compressive} & 2.98431 & 4.05576 & 10.44520 & 126.424987\\
 \text{10$\%$ Tensile} & 3.64749 & 4.95704 & 10.44520 & 188.857084\\
  \text{4$\%$ Compressive} & 3.18326 & 4.32614 & 10.44520 & 143.843231\\
 \text{4$\%$ Tensile} & 3.44854 & 4.68666 & 10.44520 & 168.816718\\
 \underline{Out-of-plane}\\
 \text{10$\%$ Compressive} & 3.31590 & 4.50640 & 9.40065 & 140.471775\\
 \text{10$\%$ Tensile} & 3.31590 & 4.50640 & 11.48970 & 171.687978\\
  \text{4$\%$ Compressive} & 3.31590 & 4.50640 & 10.02740 & 149.837157\\
 \text{4$\%$ Tensile} & 3.31590 & 4.50640 & 10.86300 & 162.323337\\
 \\
~& $\alpha$ (\text{deg}) & $\beta$ (\text{deg}) & $\gamma$ (\text{deg}) & ~ \\
  ~& $90^{\circ}$  & $90^{\circ}$  & $90^{\circ}$  & ~ 
  \end{tabular}
  }
\end{ruledtabular}
\end{table}
\endgroup

\begin{figure*}
{\includegraphics[width=6.2cm,height=4.60cm]{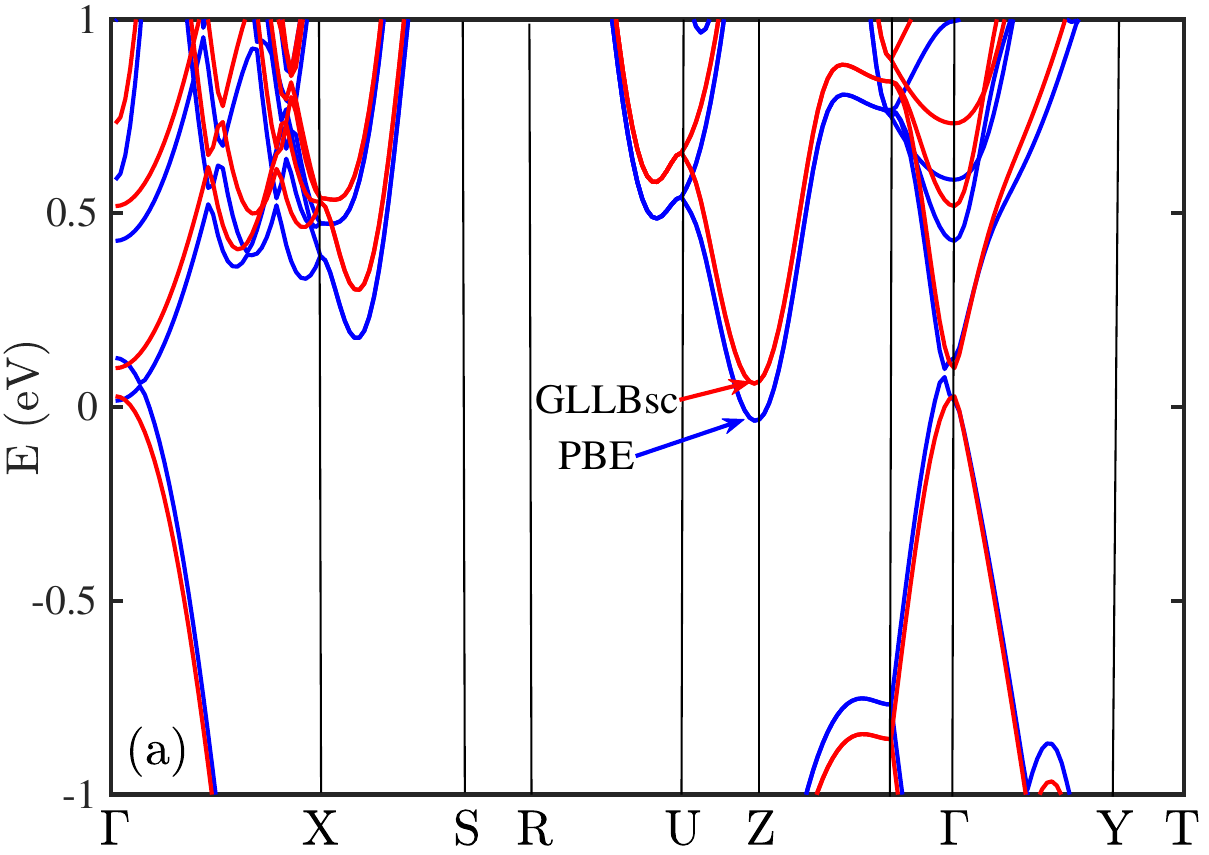}
\includegraphics[width=11.50cm,height=4.30cm]{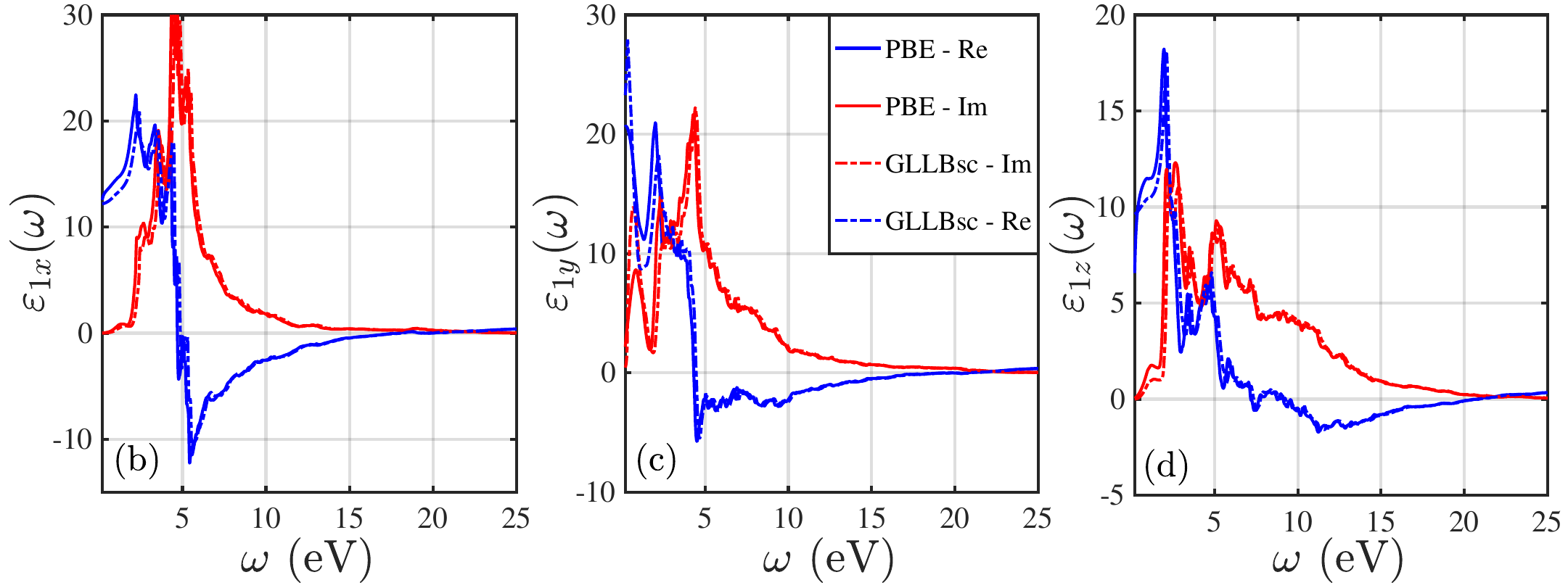}}
\caption{
(Color online). Band structure and permittivity components for unstrained BP using PBE and GLLBsc functionals. (a) The band structure is shown for PBE and GLLBsc by the blue and red curves, respectively. (b)-(c) Plots of the permittivity tensor components ($\varepsilon_{1x,1y,1z}$) as a function frequency. The solid and dashed curves show PBE and GLLBsc results, respectively. The real (Re) and imaginary (Im) parts are displayed by the blue and red colors, respectively.
}
\label{pbe-gllbsc}
\end{figure*}

\section{DFT simulations of unit cell parameters, electronic band structure, and density of states} \label{appendix2}

The lattice parameters ($a_0, b_0, c_0, \alpha, \beta, \gamma$) and the location of AB stacking order for
phosphorus atoms ($x,y,z$) 
in an unstrained unit cell are summarized in Table \ref{abctab}. 
To simulate strained BP, we have scaled both the 
unit cell and atomic locations (in percent, with respect to the unstrained parameters), depending on the type 
of strain applied. To characterize the material response to strain, we also list the in-plane covalent bonds ($\ell_1$),
  interlayer distance ($\ell_2$), and in-plane dihedral angle ($\Gamma$) for tensile and compressive $10\%$ 
  in-plane strains:
\begin{equation}
\begin{array}{cccc}
& \text{compressive} & \text{unstrained} & \text{tensile}\\
{\ell_1} (\text{\AA}) & 2.00521 & 2.22802  & 2.45082\\
{\ell_2} (\text{\AA}) & 2.23537 & 2.26009  & 2.28711\\
\Gamma (\text{deg}) & 27.0833 & 29.9636  & 32.8056
\end{array}.
\end{equation}
In Table \ref{abctab2}, we summarize the lattice parameters for BP when it is subject to in-plane or out-of-plane compressive and tensile strains. 
Note that the normalized positions of the atoms are identical to those given in Table \ref{abctab}.

We present in Fig.~\ref{bs} the calculations for the electronic band structure (left panels) and
total density of states (TDOS) (right panels) for bulk black phosphorus with both in-plane and 
out-of-plane compressive and tensile stresses. From the top row, it is evident that the band 
gap is sensitive to the in-plane strain, and whether it is of the compressive or tensile type.
Importantly, there are no band crossings along any of the symmetry points in Fig.~\ref{bs}(c), 
and due to the applied tensile strain a clear gap emerges at the Fermi energy. Upon
compressive strain, the interatomic spacing is reduced, the gap vanishes, and the system  
becomes conductive. Depending on the applied strain, the band structure exhibits an increasing number
of interband transitions leading to increased losses. The opposite occurs when 
tensile forces are applied, and the corresponding modified lattice constants lead to a band
gap and changing dielectric optical properties. This is consistent with the top row of
Fig.~\ref{ENZ1},  where it was shown that as BP experiences increased compressive 
strain ($\epsilon_{zz}$ decreases), there is a greater dissipation as seen in the increased 
imaginary component of the permittivity. Similar behavior is seen at these frequencies when 
there is in-plane strain (not shown).

The tunability of the band gap of BP is very well understood. \cite{bgclosing1,bgclosing2}
As can be seen in Fig.~\ref{bs}(a), the DFT-predicted band gap of bulk BP is zero while the experimental 
optical band gap is $0.3$ eV. In effect, this discrepancy originates from the underestimation
  of the BP band gap by standard DFT functionals, such as PBE. To improve the band gap prediction, one may
  either repeat the calculation by a hybrid functional or resort to the GW approximation for the contribution
  of self-energy.~\cite{GWA,PBE0,bgclosing1,bgclosing2} While GW has a more solid fundamental footing,
  DFT provides better agreement with experimental band gaps for some material systems. For example, in
  the case of single-layer $\rm MoS_2$, DFT-PBE and DFT-HSE (hybrid functional) yield reasonable values
  (1.6 eV and 1.9 eV, respectively)~\cite{dp1}, very close to the photoluminescence (PL) experimental
  value (1.8-1.9 eV). ~\cite{dp2,dp3} Conversely, the band gap of single-layer $\rm MoS_2$ is seriously
  overestimated by the GW method (2.7 eV).~\cite{dp1} It is well known that exciton effects, ~\cite{dp4}
  absent in standard GW calculations, lead to a lower effective band gap as found in PL experiments. Hence,
  in some cases DFT methods, in particular based on hybrid exchange-correlation functional approximations,
  can lead to band gaps that fit experimental results better than those for GW.
 
     \begin{figure}[t] 
\centering
\includegraphics[width=0.48\textwidth]{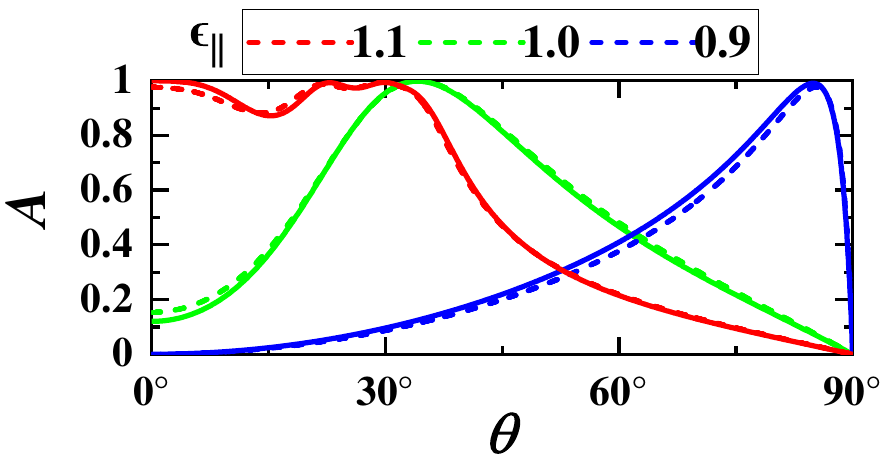}
\caption{(Color online). Absorptance ($A$) vs the incident angle $\theta$ for three in-plane strain values $\epsilon_{||}=1.1, 1.0, 0.9$. The frequency of the incident wave is set at $\omega=22.6$~eV. The corresponding permittivity components of the dashed and solid lines are obtained using the GLLBsc and PBE functionals, respectively.}
\label{compare}
\end{figure}  
  
To illustrate that the application of different functionals can improve the band gap
  underestimation by PBE while providing negligible difference in linear response, we have plotted the
  band structure and the corresponding components of the permittivity tensor for unstrained BP using both
  the PBE and GLLBsc functionals \cite{gllbsc1,gllbsc2} in Fig. \ref{pbe-gllbsc}. As is clearly seen in
  Fig. \ref{pbe-gllbsc}(a), the (indirect) band gap at the $\Gamma$ and $Z$ points opens up to $\sim 0.1$~eV
  for GLLBsc. A comparison of the permittivity tensor components in Figs. \ref{pbe-gllbsc}(c)-\ref{pbe-gllbsc}(d)
  displays only slight modifications between PBE and GLLBsc. This can be understood by the fact that the
  dielectric response is a collective response, meaning that momentum space is integrated out within the
  Brillouin zone. Therefore, slight shifts of electronic bands do not severely alter the dielectric response. Consequently, by simply looking at DOS or band structure it is impractical to make a conclusion about the behavior or the dielectric response of BP.
In any case, one should note that the random phase approximation used for calculating the dielectric
response tensor does not include exchange-correlation contributions (although 
there is surely a dependence via the pre-determined ground-state electron density).\cite{PN} We emphasize
that the presented results and conclusions made for the absorption in BP-based heterostructures rely only
on the ENZ mechanism in the low-dissipation regime. Our calculations reveal that several ENZ modes
are accessible throughout the frequency interval, and any possible band gap corrections will not affect the main message of this work. To demonstrate this fact, we have plotted in Fig.~\ref{compare}, the angle-dependent absorptance by 
using both the PBE and GLLBsc functionals. The dashed and solid curves 
correspond to the GLLBsc and PBE functionals, respectively. 
Note that all parameters are identical to those used in Fig.~\ref{a_vs_theta2}(a). As seen, the use of the different functionals results in negligible variations in absorptance.

\end{document}